\documentclass[fp,twocolumn]{jpsj3}
\usepackage{txfonts}
\usepackage{braket}
\title{Tight-Binding Model and Electronic Property of Dirac Nodal Line in Single-Component Molecular Conductor [Pt(dmdt)$_{2}$]}

\author{Taiki~Kawamura$^{1}$, Daigo~Ohki$^{1}$, Biao~Zhou$^{2}$, Akiko~Kobayashi$^{2}$, and Akito~Kobayashi$^{1}$}
\inst{$^{1}$Dept. of Phys., Nagoya Univ, Chikusa-Ku, Nagoya city, Aichi 464-8602, Japan\\
$^{2}$Dept. of Chem., College of Humanities and Sciences, Nihon Univ, Setagaya-Ku, Tokyo 156-8550, Japan}

\abst{Motivated by the recent discovery of Dirac nodal line in the single-component molecular conductor
[Pt(dmdt)$_{2}$], we propose a three-orbital tight-binding model based on the Wannier fitting of the first-principles calculation, and address the problems of edge states, topological properties and magnetic susceptibility. We find that logarithmic peaks of the local density of states emerge near the Fermi energy, owing to pseudo-one-dimensional edge states that appear between the Dirac nodal lines. 
Magnetic susceptibility calculated in our model can explain the experimental result at a high temperature. In the presence of a realistic spin-orbit coupling, we show that [Pt(dmdt)$_{2}$] is a topological nodal line semimetal with isolated electron and hole pockets. 
}

\usepackage{color}

\begin{document}
\maketitle

\section{Introduction}
The Dirac electron system in condensed matter has been found in the energy dispersion of graphite\cite{Wallace1947,McClure1957} and many other materials has attracted attention such as bismuth\cite{Wolff1964}, 
graphene\cite{Novoselov2005}, and the organic conductor $\alpha$-(BEDT-TTF)$_{2}$I$_{3}$\cite{Kajita1992,Tajima2000,Kobayashi2004,Katayama2006,Kobayashi2007,Goerbig2008}.
Graphene and $\alpha$-(BEDT-TTF)$_{2}$I$_{3}$ have two-dimensional Dirac cones that degenerate at the Dirac points in the vicinity of the Fermi energy and exhibit similar behaviors to relativistic massless fermions.
In three-dimensional momentum space, Dirac nodal lines are drawn by connecting Dirac points
\cite{Burkov2011,Chiu2014,Fang2015,Gao2016}.
Graphite is a three-dimensional Dirac semimetal whose electron and hole pockets are connected along the Dirac nodal lines, since graphite has a layered structure of graphene sheets.
Recently, in addition to graphite, various Dirac nodal line semimetals have been found; such semimetals are transition-metal monophosphates\cite{Weng2015}, Cu$_3$N\cite{Kim2015}, antiperovskites\cite{Yu2015}, perovskite iridates\cite{Carter2012}, 
the hexagonal pnictides CaAgX (X = P, As)\cite{Yamakage2016}, 
and the single-component molecular conductors [Pd(dddt)$_2$]
\cite{Kato2017JACS,Kato2017JPSJ,Suzumura2017JPSJ,Suzumura2017JJAP,SuzumuraKato2018,SuzumuraYamakage2018,Tsumuraya2018,Suzumura2019}
and [Pt(dmdt)$_{2}$]\cite{Zhou2019}.
The Dirac electron systems exhibit anomalous behaviors in large orbital diamagnetism\cite{Fukuyama1970,Fuseya2015}, electrical resistivity\cite{Ando2005}, Hall coefficient\cite{Fukuyama2007,Kobayashi2008,Tajima2010}, 
and topological edge states\cite{Kohmoto2007}.
In the Dirac nodal line semimetals, a flat Landau level\cite{Rhim2015}, the Kondo effect\cite{Mitchell2015}, long-range Coulomb interaction\cite{Huh2016}, and a quasi-topological electromagnetic response\cite{Ramamurthy2017} are also expected.
The Dirac nodal line semimetals have become a hot topic.

It has been shown that the electrical resistivity of [Pt(dmdt)$_{2}]$ is almost independent of the temperature $T$, and the magnetic susceptibility excluding the magnetic impurity effect decreases as $T$ decreases\cite{Zhou2019}.
Those are basic features of an ideal Dirac electron system where Dirac points or nodal lines are located near the Fermi energy\cite{Kajita2014}.
Dirac nodal lines have been found in the energy dispersion obtained by first-principles calculation\cite{Zhou2019}.
Recently, a two-orbital tight-binding model based on the extended H\"{u}ckel method has been proposed, showing that this model has Dirac nodal lines\cite{Kato2020}.
However, a model of [Pt(dmdt)$_{2}$] based on first-principles calculation has not been elucidated yet.

In the present paper, we construct a three-orbital tight-binding model based on the Wannier fitting of the first-principles calculation, focusing on three isolated bands near the Fermi energy.
We confirm that the energy dispersion, Fermi surfaces, and density of states (DOS) obtained by first-principles calculation are reproduced by the three-orbital tight-binding model.
We also calculate the Berry curvature to elucidate the unique properties of the Dirac electron system.
Furthermore, we investigate the effects of spin-orbit coupling (SOC) on the Fermi surfaces and edge states.
We find that [Pt(dmdt)$_{2}$] is a topological nodal line semimetal with isolated electron and hole pockets in the presence of a realistic weak SOC, where the local density of states (LDOS) of the edge states has logarithmic peaks near the Fermi energy.
As SOC increases, we find a qualitative change of the Fermi surface: a crossover to a weak topological insulator with pseudo-one-dimensional helical edge states. 
In addition, we calculate the magnetic susceptibility. 
We show that the $T$-linear behavior of the magnetic susceptibility at a high temperature is consistent with the experimental result\cite{Zhou2019}, supporting the existence of Dirac nodal line.
We argue that the depletion of the magnetic susceptibility observed experimentally at a low temperature\cite{Zhou2019} cannot be explained only by the effect of a realistic SOC.

The remainder of this paper is organized as follows.
In Sect. 2, we present a tight-binding model based on the Wannier fitting of the first-principles calculation of [Pt(dmdt)$_{2}$].
In Sect. 3, we confirm that the present model reproduces the isolated three bands near the Fermi energy obtained by first-principles calculation. 
In Sect. 4, we present the spectrum, DOS, and energy dispersion of edge states.
In Sect. 5, we also present the effects of SOC on the bulk and edge states.
In Sect. 6, the topological properties of [Pt(dmdt)$_{2}$] with the bulk-edge correspondence are shown.
In Sect. 7, the magnetic susceptibility in the present model is shown. 
In Sec. 8, a summary and discussion are given.
In Appendix A, we focus on two-dimensional simplified models and confirm the stability of Dirac points.
In Appendix B, we show the correspondence between the present model and the two-orbital model obtained by the extended H\"{u}ckel method\cite{Kato2020}.

\section{Tight-Binding Model Based on Wannier Fitting}

The crystal structure of [Pt(dmdt)$_{2}$] is shown in Fig. 1\cite{Zhou2019}.
This material is composed of single-component molecules and has three-dimensionality.
Each side of the unit cell is composed of lines connecting Pt atoms and has spatial inversion symmetry centered on a Pt atom.
The crystal structure of [Pt(dmdt)$_{2}$] is a triclinic system. 

\begin{figure}[htpb]
\begin{center}
\includegraphics[width=50mm]{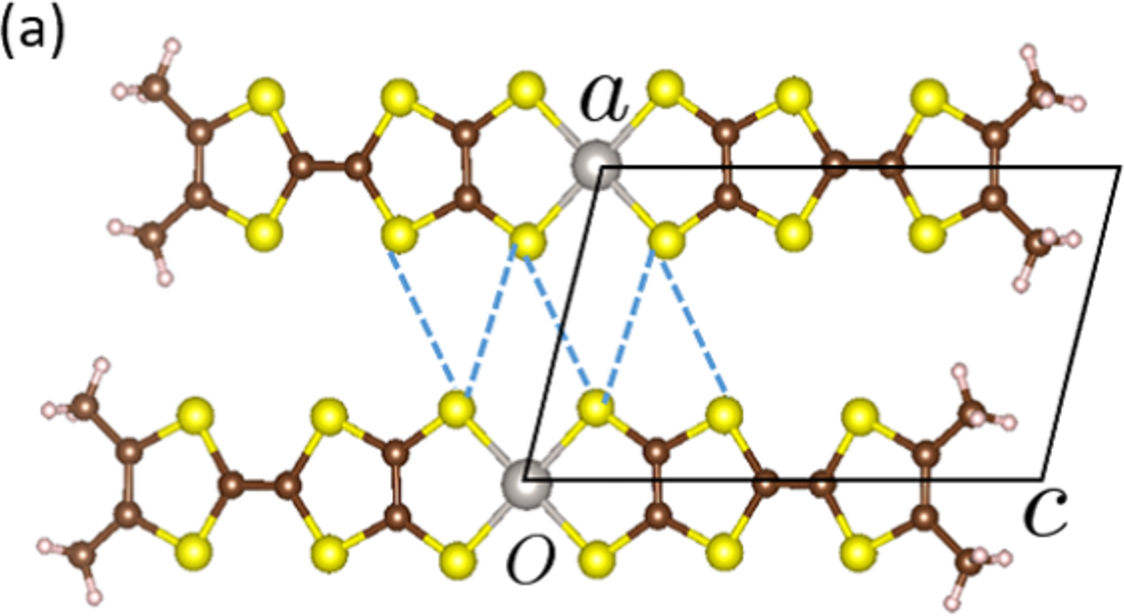}
\includegraphics[width=60mm]{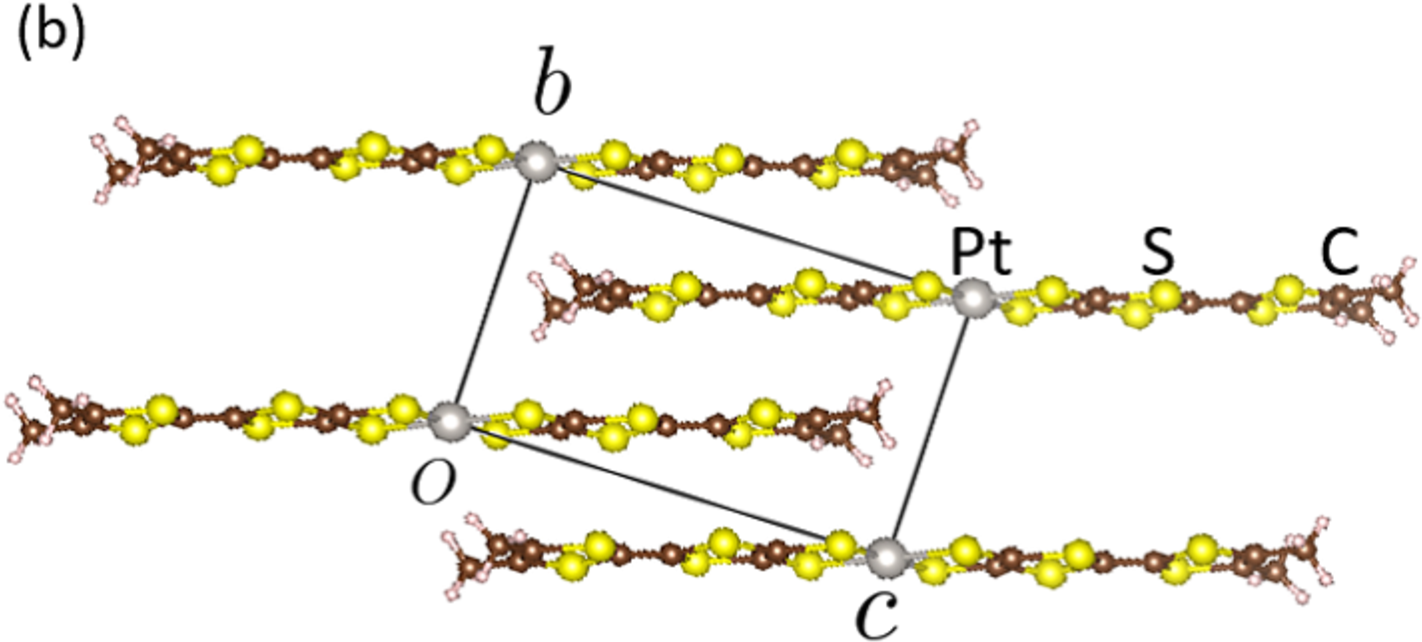}
\end{center}
\caption{(Color online) Crystal structure of the single-component molecular conductor [Pt(dmdt)$_{2}$] viewed from the (a) vertical axis and (b) side of the molecule. The solid line shows a unit cell. Blue dashed lines denote short S $\cdots$ S contacts.}
\label{f1}
\end{figure}

We calculate the energy band structure of [Pt(dmdt)$_{2}$] using Quantum ESPRESSO\cite{Giannozzi2017} on the basis of the results of X-ray structure analysis in order to obtain hopping energies using RESPACK\cite{Nakamura2020}. 
In the previous paper\cite{Zhou2019}, the first-principles calculation was carried out using QMAS code\cite{Ishibashi2007}.
Quantum ESPRESSO and QMAS code are packages for first-principles calculation.
Quantum ESPRESSO is based on the pseudopotential method, where we use the norm-conserving pseudopotential, while QMAS code is based on the PAW method. The latter is more accurate than the former, but the results of both calculations are consistent in the present case.
Figure 2(a) shows the energy band structure calculated by first-principles calculation. 
The horizontal axis connects the symmetry points in the first Brillouin zone (BZ) and the vertical axis is the energy measured from the Fermi energy. 
A characteristic of the energy band structure of [Pt(dmdt)$_{2}$] is the three isolated bands near the Fermi energy\cite{Zhou2019}.
We construct the three-orbital tight-binding model focusing on these three isolated bands by Wannier fitting.
Figure 2(b) shows the energy band structures obtained by first-principles calculation reproduced accurately by the three-orbital tight-binding model including the hopping energies and site potentials.
Figures 2(c) and 2(d) shows the molecular structure of [Pt(dmdt)$_{2}$] and the three Wannier orbits (sites 1, 2, and 3) constituting the present three-orbital tight-binding model.
In Fig. 2(c), the Pt(dmdt)$_{2}$ molecule and the Wannier orbits viewed from the vertical axis of the molecule are drawn. 
In Fig. 2(d), the Wannier orbits viewed from the side of the molecule are drawn, where the navy line and red dot express a Pt(dmdt)$_{2}$ molecule and a Pt atom, respectively. 
These Wannier orbits are distributed throughout the molecule. 
Site 1 is centered on the Pt atom, while sites 2 and 3 are unevenly distributed toward one side.

\begin{figure}[htpb]
\begin{center}
\includegraphics[width=60mm]{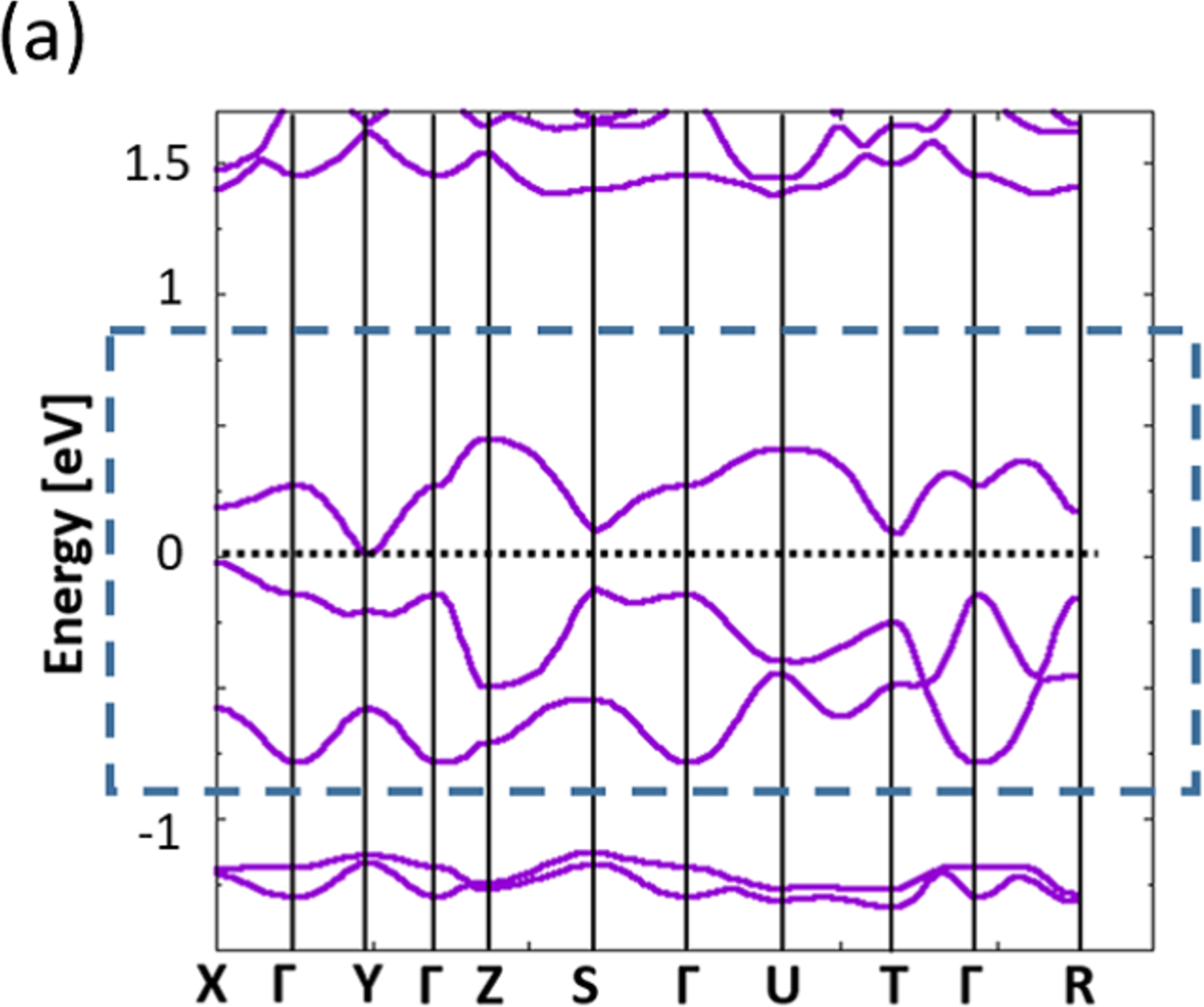}
\includegraphics[width=60mm]{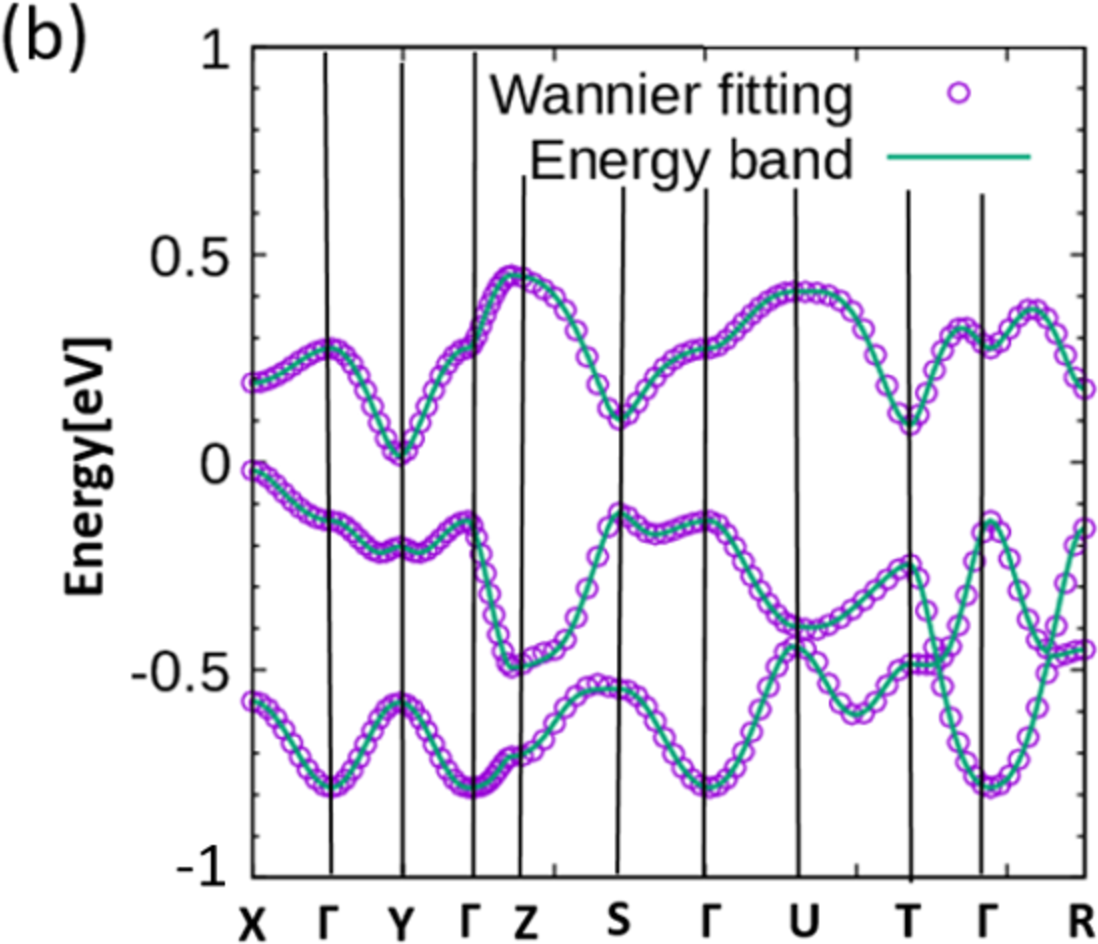}
\includegraphics[width=60mm]{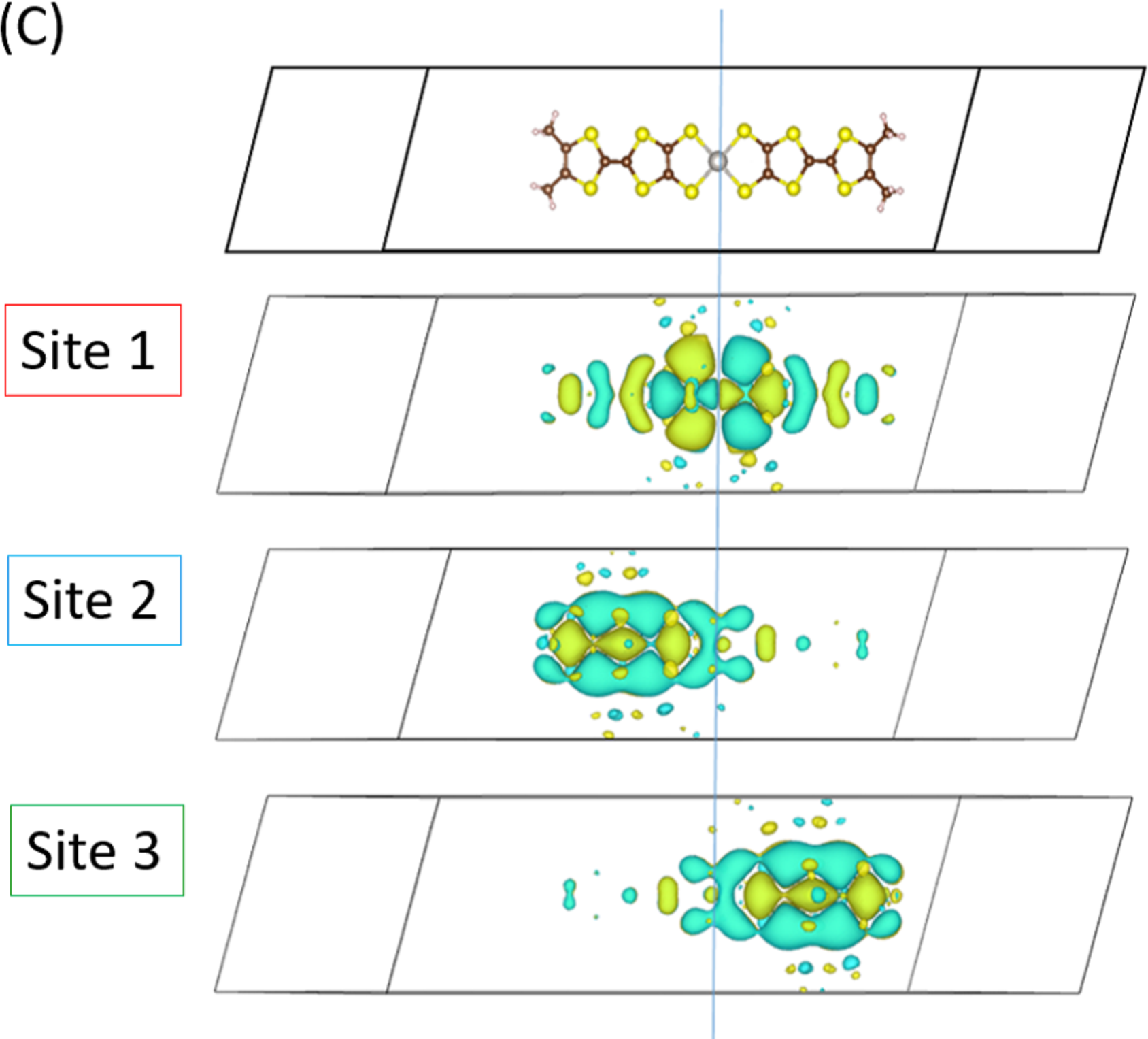}
\includegraphics[width=60mm]{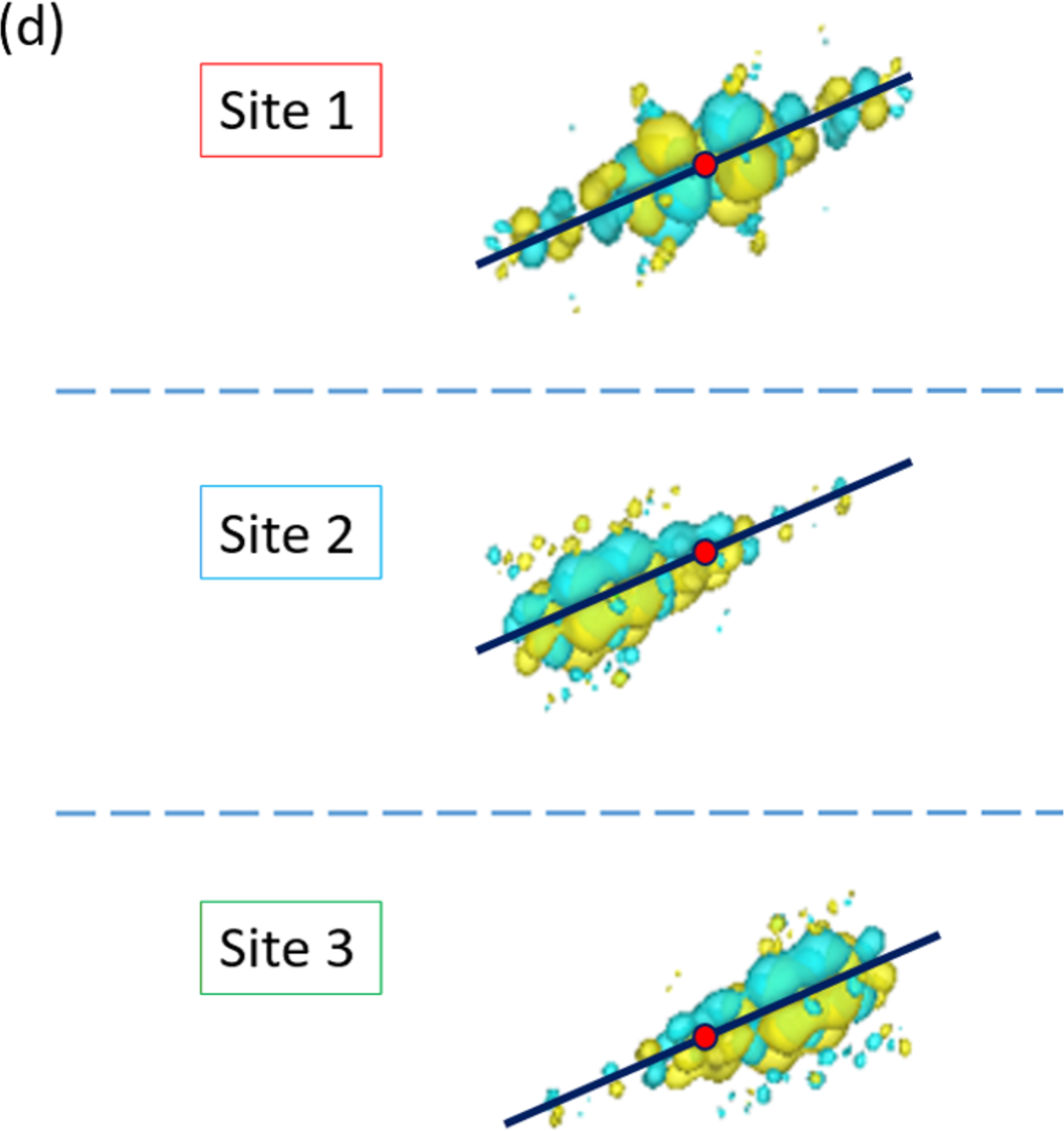}
\end{center}
\caption{(Color online)
(a) The energy band structure of [Pt (dmdt)$_{2}$] calculated by first-principles calculation. The horizontal axis connects the symmetry points in the first BZ. Three bands near the Fermi energy (black dotted line) are isolated from the other bands\cite{Zhou2019}.
(b) Energy band structures obtained by first-principles calculation (green lines) and the present three-orbital tight-binding model (purple circles). 
(c) Pt(dmdt)$_{2}$ molecule and Wannier orbits (sites 1, 2, and 3) viewed from the vertical axis of the molecule.
(d) Wannier orbits viewed from the side of the molecule. The navy bold line and the red central dot express a Pt(dmdt)$_{2}$ molecule and Pt atom, respectively.}
\label{f1}
\end{figure}

The three-orbital tight-binding model is given by 
\begin{equation}
H=\sum_{i,j,\alpha,\beta,\sigma}t_{i\alpha;j\beta,\sigma}c^{\dagger }_{i\alpha\sigma} c_{j\beta\sigma}~,
\end{equation}
where $i$ and $j$ represent unit cells and, $\alpha$ and $\beta$ represent Wannier orbits (sites 1, 2, and 3). 
$\sigma$ is the spin index. 
$t_{i\alpha;j\beta,\sigma}$ for the two different sites represents the hopping energy 
and $t_{i\alpha;j\beta,\sigma}$ for the same site ($i=j$ and $\alpha =\beta$) represents the site potential. 
By Fourier transformation, we obtain the Hamiltonian in the wavenumber picture.
\begin{equation}
H(\textbf{k})=\sum_{\textbf{k},\alpha,\beta,\sigma} H_{\alpha\beta\sigma}(\textbf{k}) c^{\dagger}_{\textbf{k}\alpha\sigma}c_{\textbf{k}\beta\sigma}
\end{equation}
Hereafter, the energies are given in eV.
In the present paper, we take all hopping energies and site potentials whose absolute values are larger than $0.01$.
The networks of the hopping energies are drawn in Figs. 3(a) and 3(b). 
The hopping energies in the $b$-$c$ plane are shown in Fig. 3(a), and the hopping energies that include hopping energies out of the $b$-$c$ plane are shown in Fig. 3(b).
The hopping energies obtained by the Wannier fitting are given as $t_{1}=0.212$, $t_{2}=0.179$, $t_{3}=0.201$, $t_{4}=0.044$, $t_{5}=0.043$, $t_{6}=0.042$, $t_{7}=0.014$, $t_{8}=0.024$, $t_{9}=0.013$, $t_{10}=0.011$, $t_{11}=0.051$, and $t_{12}=0.051$. 
The site potential of site 1 obtained by the Wannier fitting is $\Delta =0.07$ larger than those of the other two sites.
In addition, we treat the SOC\cite{Kane-Mele,OsadaET,OsadaTau}.
The hopping energy $t$ is modified as $t(1+i\nu\lambda \sigma)$,
where $\nu =\pm1$ is proportional to the outer product of momenta and the gradient of potential energy, $\textbf{P}\times\nabla U$. 
$\lambda$ is the SOC constant, which is treated as a parameter in the present paper.
We carried out this modification for the hopping energies whose absolute values are larger than 0.04, while most hopping energies were not modified because of the hopping direction and surrounding sites.
The matrix elements of the Hamiltonian are given by

\begin{eqnarray}
&&H_{11\sigma}=\Delta-2t_{9}\cos{k_{a}}+2t_{7}\cos{(k_{b}+k_{c})}-2t_{7}\cos{k_{c}} \\ \nonumber
&&H_{12\sigma}=t_{10}e^{i(k_{a}-k_{b})}-t_{1}e^{ik_{c}}+t_{4}(1+i\lambda \sigma)+t_{5}(1-i\lambda \sigma)e^{-ik_{b}} \\ \nonumber
&&H_{13\sigma}=-t_{5}(1-i\lambda \sigma)e^{ik_{c}}-t_{4}(1+i\lambda \sigma)e^{i(-k_{b}+k_{c})} \\ \nonumber
&&+t_{1}e^{-ik_{b}}-t_{10}e^{i(-k_{a}+k_{c})} \\ \nonumber
&&H_{22\sigma}=-2t_{8}\cos{k_{a}} \\ \nonumber
&&H_{23\sigma}=t_{11}e^{i(k_{a}-k_{b})}+t_{2}+t_{3}e^{-ik_{b}}+t_{6}e^{-i(k_{b}+k_{c})}+t_{12}e^{-ik_{a}} \\ \nonumber
&&H_{33\sigma}=-2t_{8}\cos{k_{a}}~, 
\end{eqnarray}
where $k_{a}$, $k_{b}$, and $k_{c}$ are components of the wavenumber $\textbf{k}$, $\textbf{k}=(k_{a}, k_{b}, k_{c})$. 
We define all lattice constants as 1. 
The Hamiltonian satisfies the energy eigenvalue equation
\begin{equation}
H_{\sigma}(\textbf{k})\ket{{\textbf{k},n,\sigma}}=E_{\textbf{k},n,\sigma}\ket{{\textbf{k},n,\sigma}}
\end{equation}

\[\ket{{\textbf{k},n,\sigma}}=
\left(
\begin{array}{c}
d_{1,\textbf{k},n,\sigma} \\
d_{2,\textbf{k},n,\sigma} \\
d_{3,\textbf{k},n,\sigma}
\end{array}
\right)~,
\]
where $\ket{{\textbf{k},n,\sigma}}$ is the eigenvector and $E_{\textbf{k},n,\sigma}$ is the energy eigenvalue 
of band $n$.
$d_{\alpha,\textbf{k},n,\sigma}$ is wave function of site $\alpha$. 
In the present paper, $2/3$ of the energy band is filled. Therefore, the chemical potential $\mu$ is determined by 

\begin{equation}
\frac{1}{N_{a}N_{b}N_{c}}\sum_{\text{k},n,\sigma} f_{\textbf{k},n,\sigma}=4~,
\end{equation}
where $f_{\textbf{k},n,\sigma}$ is the Fermi distribution function. $N_{a}$, $N_{b}$, and $N_{c}$ are the numbers of unit cells along $a$, $b$ and $c$ directions, respectively. Here, we define the energy eigenvalue measured from the chemical potential, $\epsilon_{\textbf{k},n}$.
\begin{equation}
\epsilon_{\textbf{k},n} \equiv E_{\textbf{k},n}-\mu
\end{equation}

\begin{figure}[htpb]
\begin{center}
\includegraphics[width=80mm]{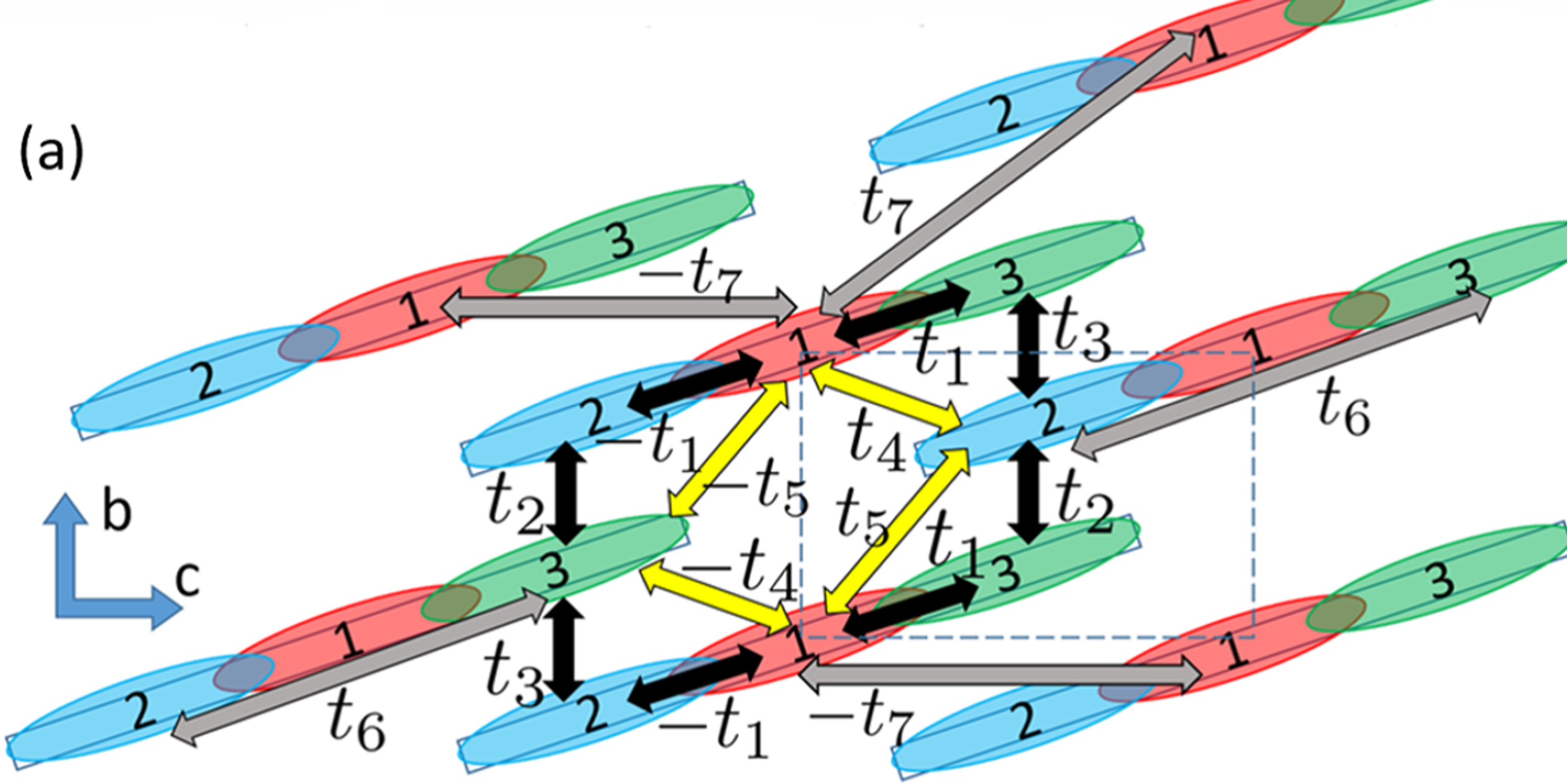}
\includegraphics[width=80mm]{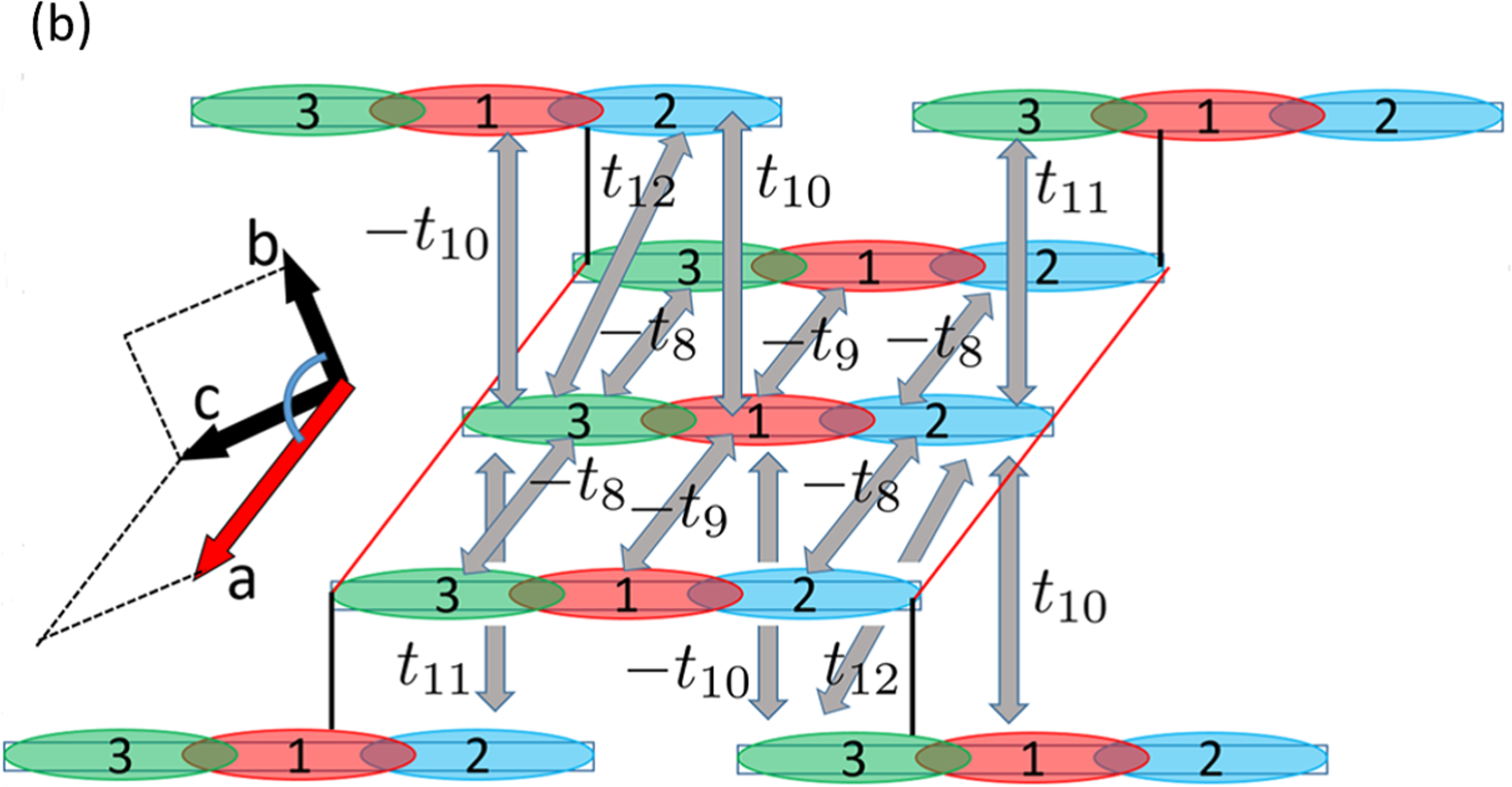}
\end{center}
\caption{(Color online)
(a) Network of the hopping energies in the $b$-$c$ plane. 
The broken box is a unit cell and the solid rectangles represent Pt(dmdt)$_{2}$ molecules.
The black arrows ($t_{1}$, $t_{2}$, and $t_{3}$) are the hopping energies whose absolute value is larger than 0.1 eV.
The yellow arrows ($t_{4}$ and $t_{5}$) are the hopping energies modified by the SOC. 
The other hopping energies are represented by the gray arrows. 
Not all hopping energies that we used are drawn in these figures for visibility. 
(b) Network of the hopping energies that include hopping along the $a$ direction. 
The red lines and black lines are additional lines to guide the eye. The red lines are parallel to the $a$ direction. The black lines connect the molecules on the same $b$-$c$ plane. 
}
\label{f1}
\end{figure}

\section{Bulk Electronic States}

We show the bulk electronic states of the present three-orbital model [Eqs. (2) and (3)] in the absence of SOC ($\lambda =0$).
Figure 4(a) shows the energy dispersion in the first BZ at $k_{a}=-\pi/2$.
Three energy bands (bands 1, 2, and 3) correspond to $\epsilon_{{\bf k}, n=1, \sigma} \ge \epsilon_{{\bf k}, n=2, \sigma} \ge \epsilon_{{\bf k}, n=3, \sigma}$ for a given ${\bf k}$. 
There are gapless linear dispersions near wavenumbers $(k_{b}/\pi,k_{c}/\pi)=(0.56, -0.28), (-0.88, 0.40)$, which are called Dirac points, between bands 1and 2. 
This result reproduces the results of first-principles calculations in both the previous paper\cite{Zhou2019} and present study. 
The Dirac points depend on $k_{a}$ and their trajectory forms the Dirac nodal lines in three-dimensional momentum space. 
Figure 4(b) shows the DOS given by $D(E)=\frac{1}{N_{a}N_{b}N_{c}}\sum_{\textbf{k},n,\sigma}\delta(E-\epsilon_{\textbf{k},n,\sigma})$, which contains the bands 1, 2, and 3.
The two peaks of DOS near -0.5 eV result from independent Van Hove singularities formed by saddle points of band 3 in the $k_{b}$-$k_{c}$ plane.
The DOS at the Fermi energy is not zero, since the Dirac points move up and down across the Fermi energy as $k_{a}$ varies.
This result also reproduces the result of first-principles calculation\cite{Zhou2019}.
Figure 4(c) shows the Dirac nodal lines formed by Dirac points between the bands 1 and 2 in the first BZ. 
The Dirac nodal lines are protected by the space inversion symmetry as discussed in Appendix A.
In addition, the Dirac nodal ring in the form of a closed loop of the Dirac nodal line exists between bands 2 and 3 as shown in Fig. 4(d), although the bands 2 and 3 overlap and the Dirac nodal ring is below the Fermi energy.

\begin{figure}[htpb]
\begin{center}
\includegraphics[width=80mm]{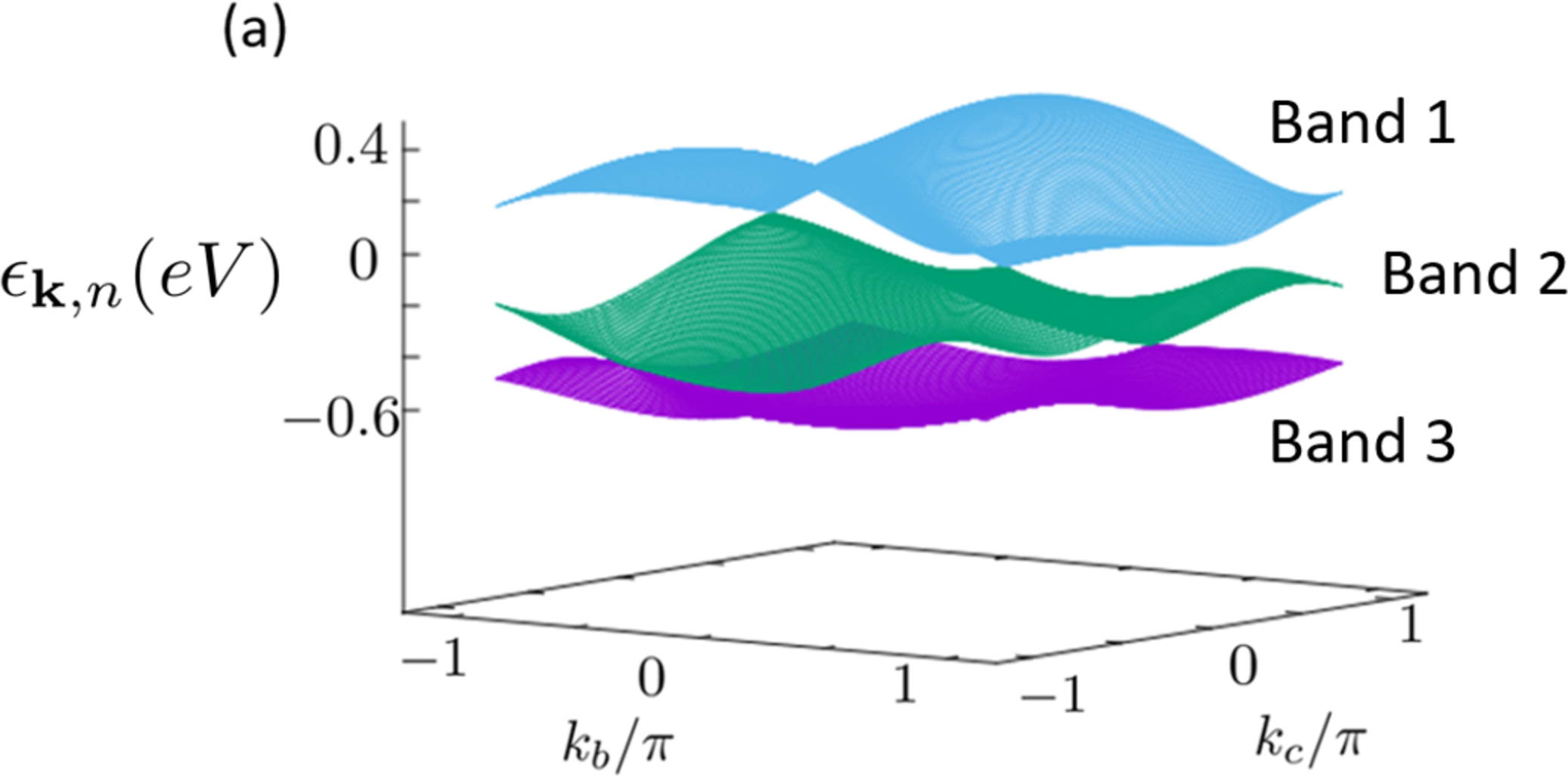}
\includegraphics[width=45mm]{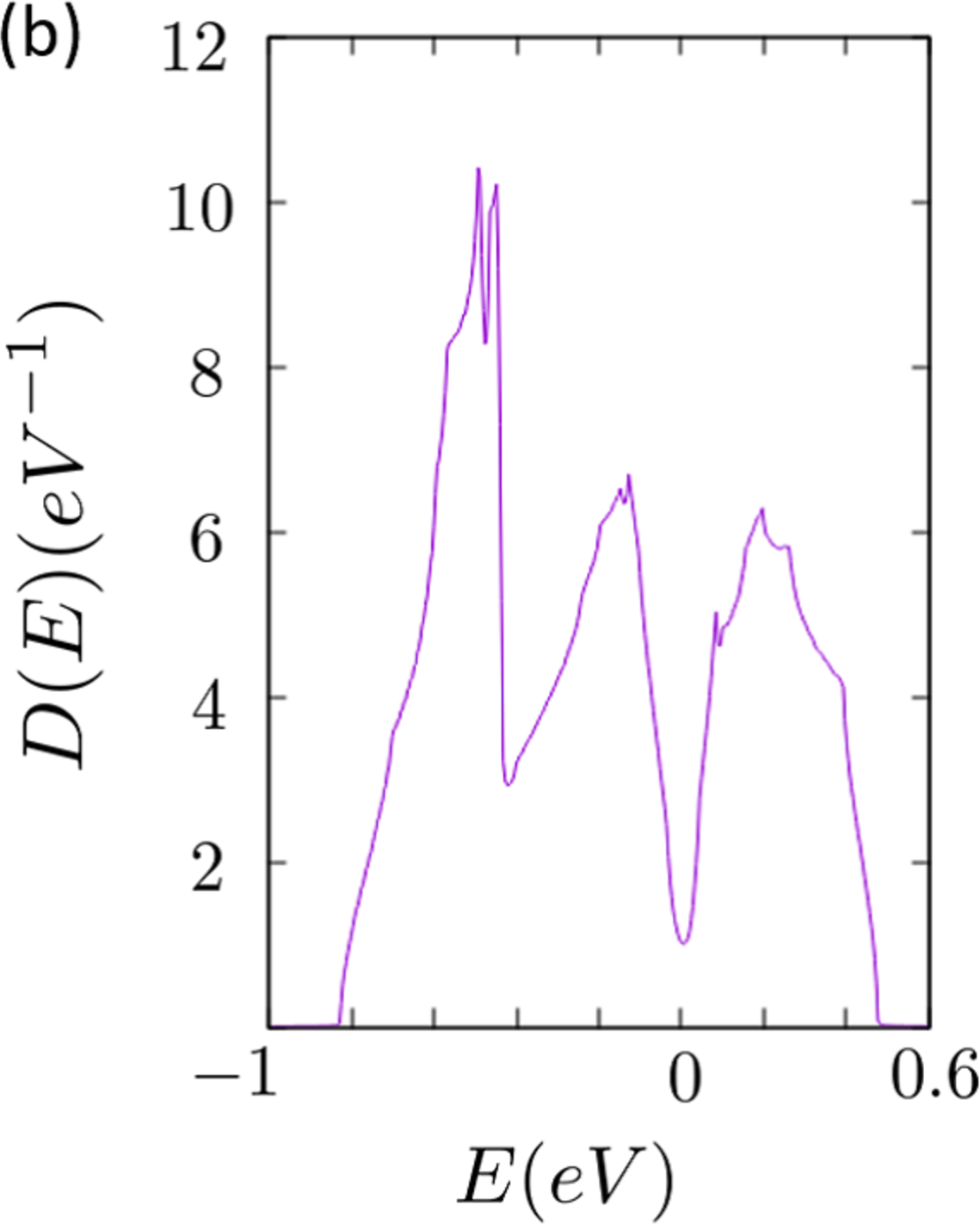}
\includegraphics[width=60mm]{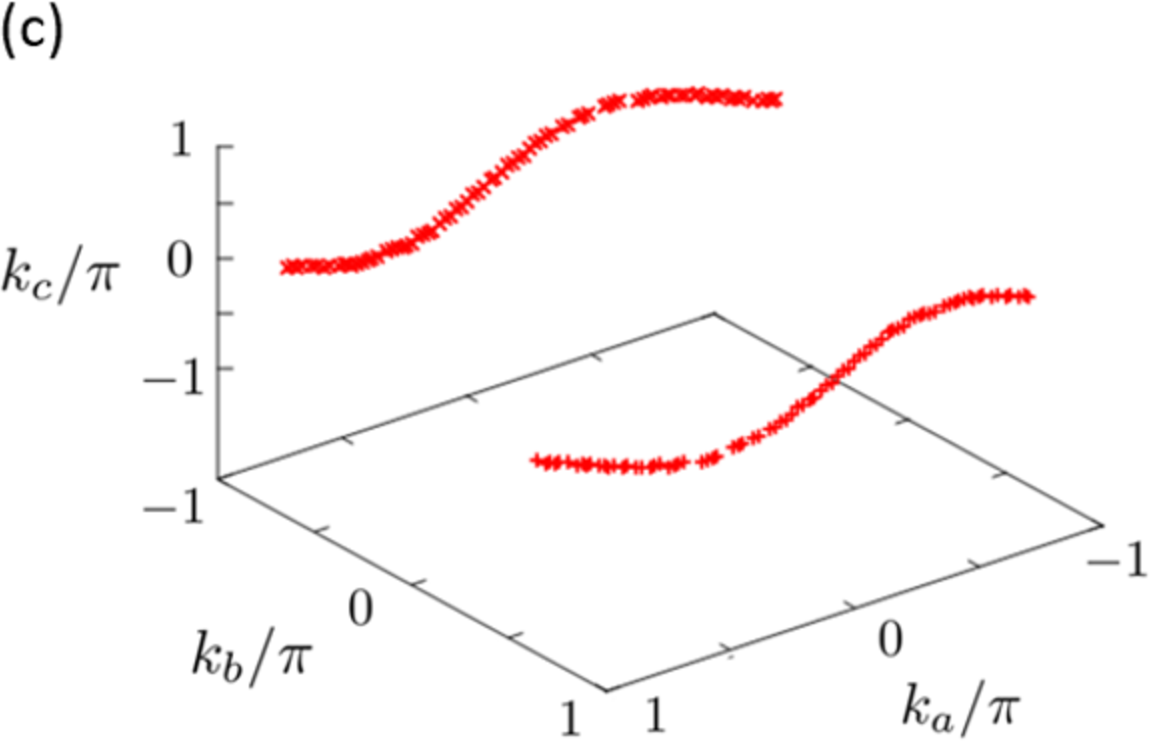}
\includegraphics[width=60mm]{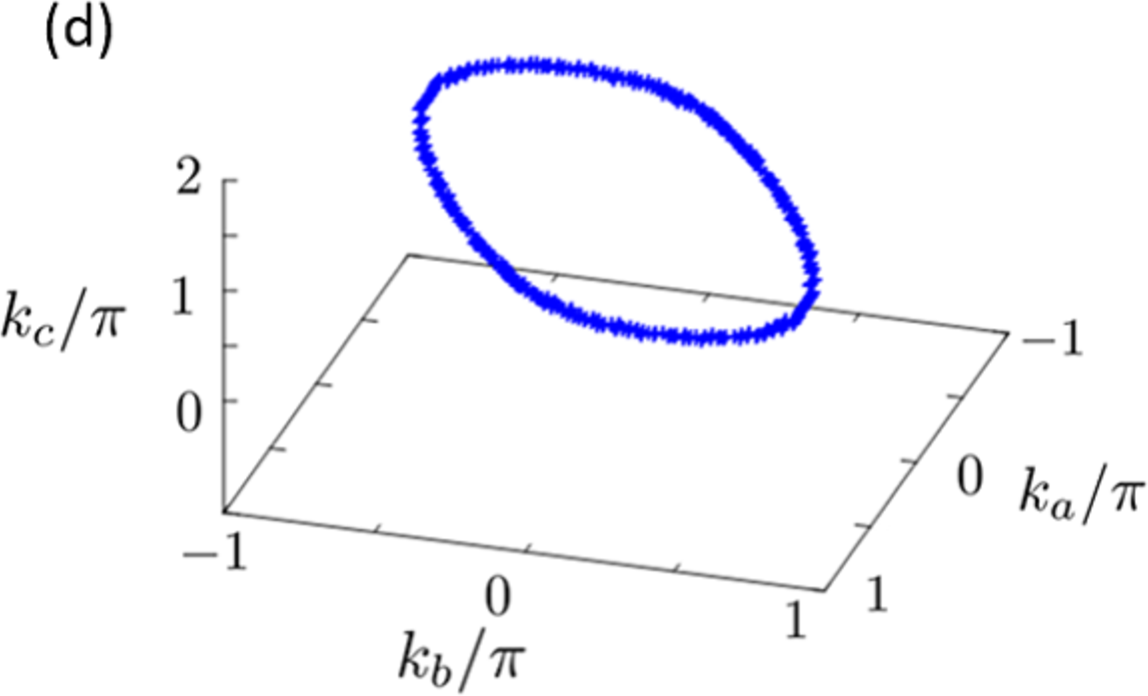}
\end{center}
\caption{(Color online)
(a) Energy dispersion of [Pt(dmdt)$_{2}$] at $k_{a}=-\pi/2$, where $k_{a}$, $k_{b}$, and $k_{c}$ are wavenumbers.
There are Dirac cones between bands 1 and 2.
(b) DOS of [Pt(dmdt)$_{2}$]. 
(c) Dirac nodal lines formed by the Dirac points between bands 1 and 2 in the first BZ.
(d) Dirac ring formed by the Dirac points between bands 2 and 3, where the plot range is not the first BZ but $-1 \le k_{a}/\pi \le1$, $-1 \le k_{b}/\pi \le1$, and $0 \le k_{c}/\pi \le 2$.}
\label{f1}
\end{figure}

Figure 5 shows the Fermi surfaces in the first BZ. 
The green Fermi surfaces are the electron pockets and the purple ones are the hole pockets.
These pockets are connected alternately along the Dirac nodal lines as shown in the previous paper\cite{Zhou2019}.
This is because the Dirac points move up and down across the Fermi energy as the wavenumber $k_{a}$ varies.

\begin{figure}[htpb]
\begin{center}
\includegraphics[width=50mm]{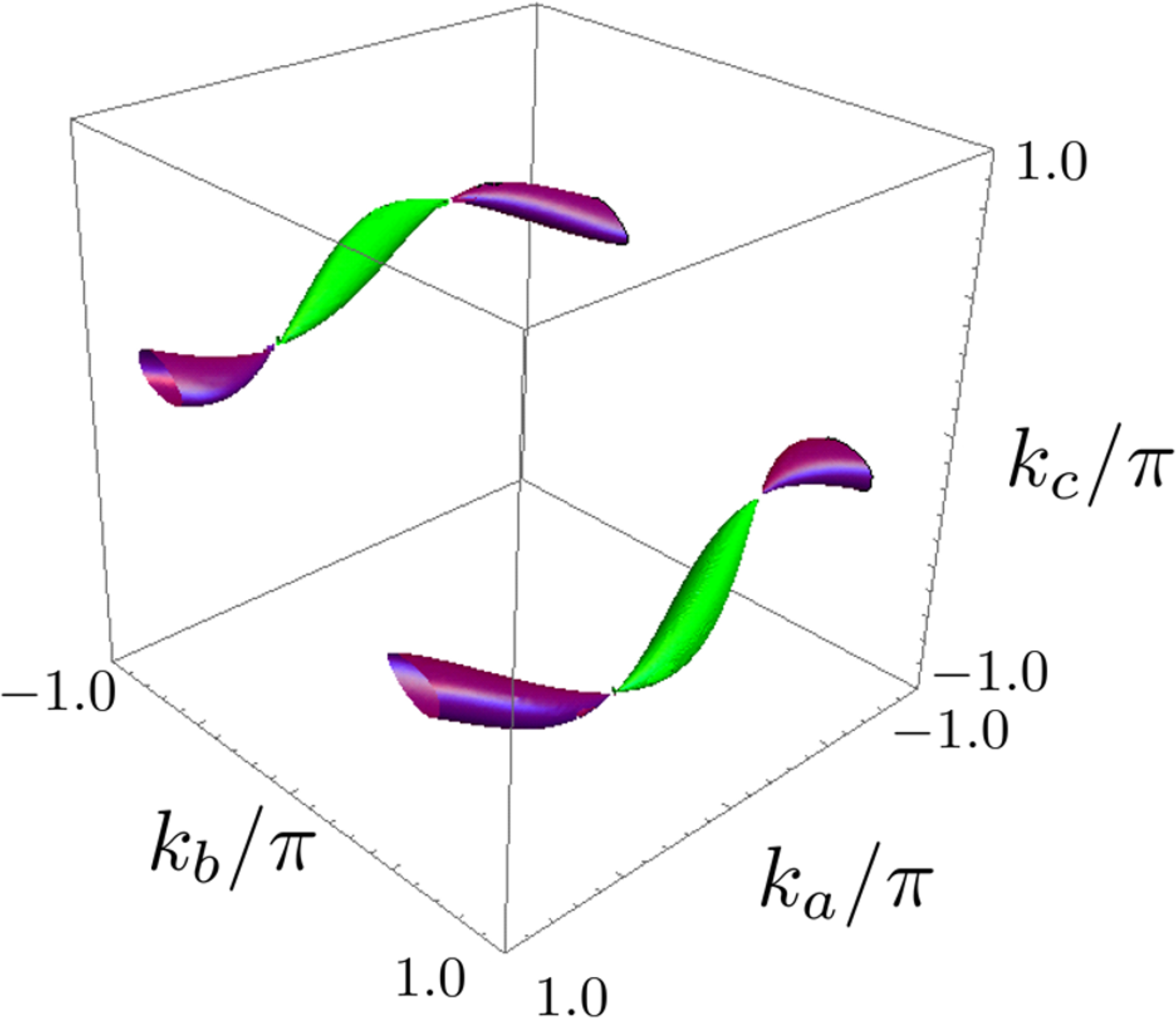}
\end{center}
\caption{(Color online)
Fermi surfaces of [Pt(dmdt)$_{2}$] in the first BZ.
The green surfaces inside BZ and the purple ones on the boundary of BZ represent the electron and hole pockets, respectively.}
\label{f1}
\end{figure}

\section{Edge States}

In the Dirac electron system, it is expected that topological edge states will appear and exhibit the bulk-edge correspondence\cite{HatsugaiPRL,HatsugaiPRB}. 
We investigate the electronic states of the present three-orbital model [Eqs. (2) and (3)] under a cylindrical boundary condition with edges perpendicular to the $c$-axis as shown in Fig. 6, where the system is periodic along the $a$ and $b$ directions.
In our calculation, treating $N_{c}$ unit cells along $c$ direction ($1 \le i_c \le N_{c}$), we solve the $3N_{c} \times 3N_{c}$ matrix Hamiltonian, where $i_c$ is the $c$-component of the unit cell coordinates $i$.
We call the edge in Fig. 6 the ``left edge" and the opposite edge the ``right edge". 
\begin{figure}[htpb]
\begin{center}
\includegraphics[width=75mm]{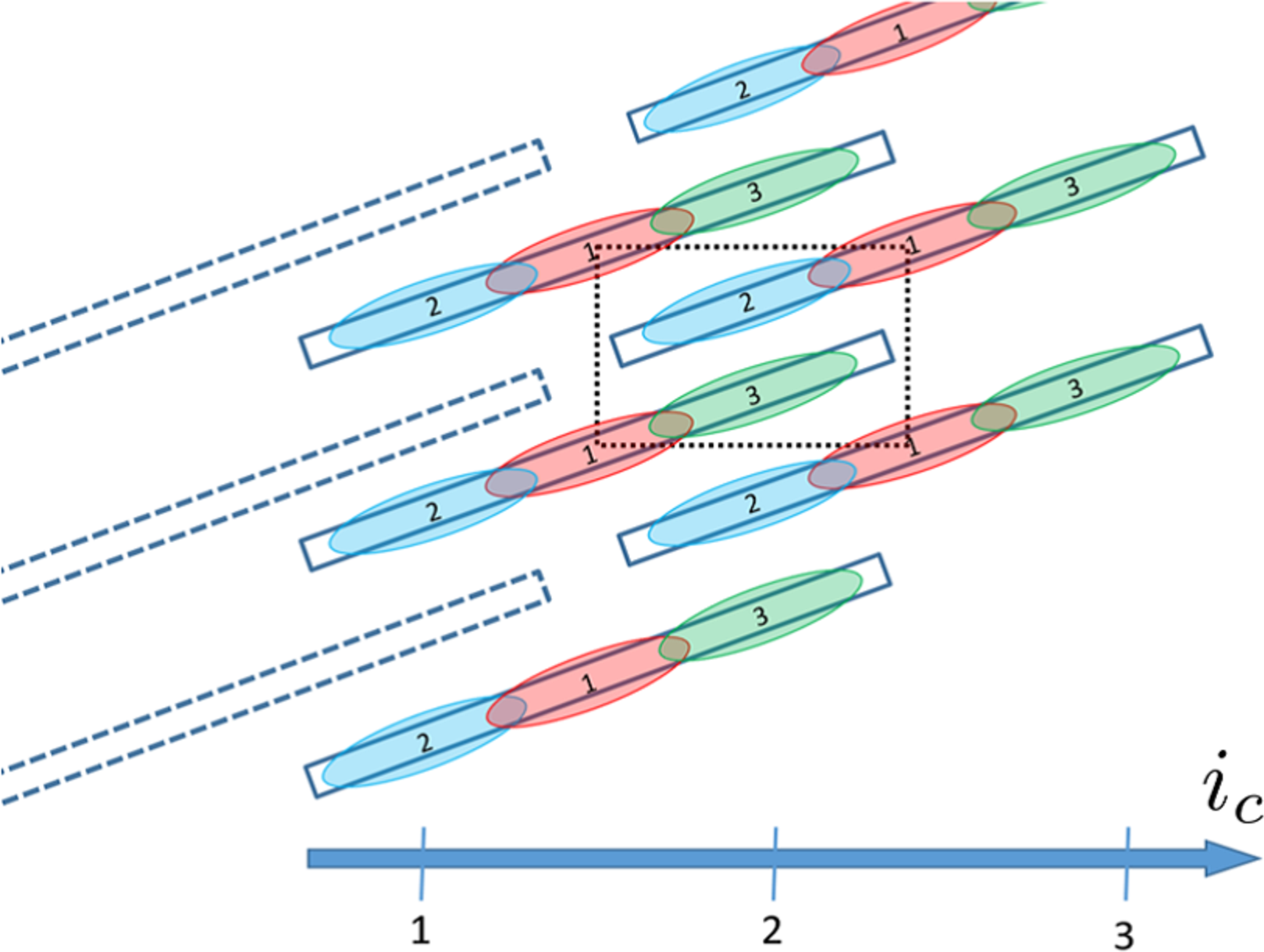}
\end{center}
\caption{(Color online)
Left edge perpendicular to the $c$-axis.
The dotted black box is a unit cell.
The broken blue rectangle shows the absence of [Pt(dmdt)$_{2}$] molecules.}
\label{f1}
\end{figure}

The site-dependent local spectral weight $A_{i ,\alpha}(\textbf{q},E)$ and the spectral weight $A(\textbf{q},E)$ are given by 

\begin{equation}
A_{i ,\alpha}(\textbf{q},E)=\frac{1}{\pi}\sum_{n,\sigma}|d_{\textbf{q},n,i,\alpha,\sigma}|^{2}
{\rm Im}\left (\frac{1}{E-\epsilon_{\textbf{q},n,\sigma}-i\eta}\right)
\end{equation}

\begin{equation}
A(\textbf{q},E)=\frac{1}{N_{c}}\sum_{i ,\alpha} A_{i ,\alpha}(\textbf{q},E)~,
\end{equation}
with $\eta =0^+$, where $d_{\textbf{q},n,i ,\alpha,\sigma}$ is the wave function with wavenumber $\textbf{q}=(q_{a},q_{b})$, band $n$, unit cell $i$, site $\alpha$, and spin $\sigma$.
$\epsilon_{\textbf{q},n,\sigma}$ is the energy eigenvalue measured from the Fermi energy. $N_{c}$ is the number of unit cells along the $c$ direction. In this calculation, $N_{a}=200$, $N_{b}=200$, and $N_{c}=40$. 

Figure 7(a) shows the spectral weight per unit cell at $q_{a}=-\pi/2$.
The horizontal axis is wavenumber $q_{b}$ and the vertical axis is the energy measured from the Fermi energy. 
We find that isolated energy eigenvalues with a flat dispersion connect two Dirac points directly and are located near the Fermi energy.
This is similar to the result for a zigzag edge in a two-dimensional honeycomb lattice\cite{Kohmoto2007}.
Figure 7(b) shows the local spectral weight of site 1 for the left edge.
We find that the isolated energy eigenvalues are localized at the edges.
Thus, they are the edge states at the $001$ surface of [Pt(dmdt)$_{2}$].
The analysis of the site-dependent local spectral weight indicates that the electrons of sites 1 and 2 localize at the left edge and the electrons of sites 1 and 3 localize at the right edge.
Figure 7(c) shows the energy dispersions that become the edge states between the two Dirac nodal lines in the $q_{a}$-$q_{b}$ plane. 
These two energy eigenvalues degenerate and show quasi-one-dimensional (quasi-1D) energy dispersions between the two Dirac nodal lines.

\begin{figure}[htpb]
\begin{center}
\includegraphics[width=70mm]{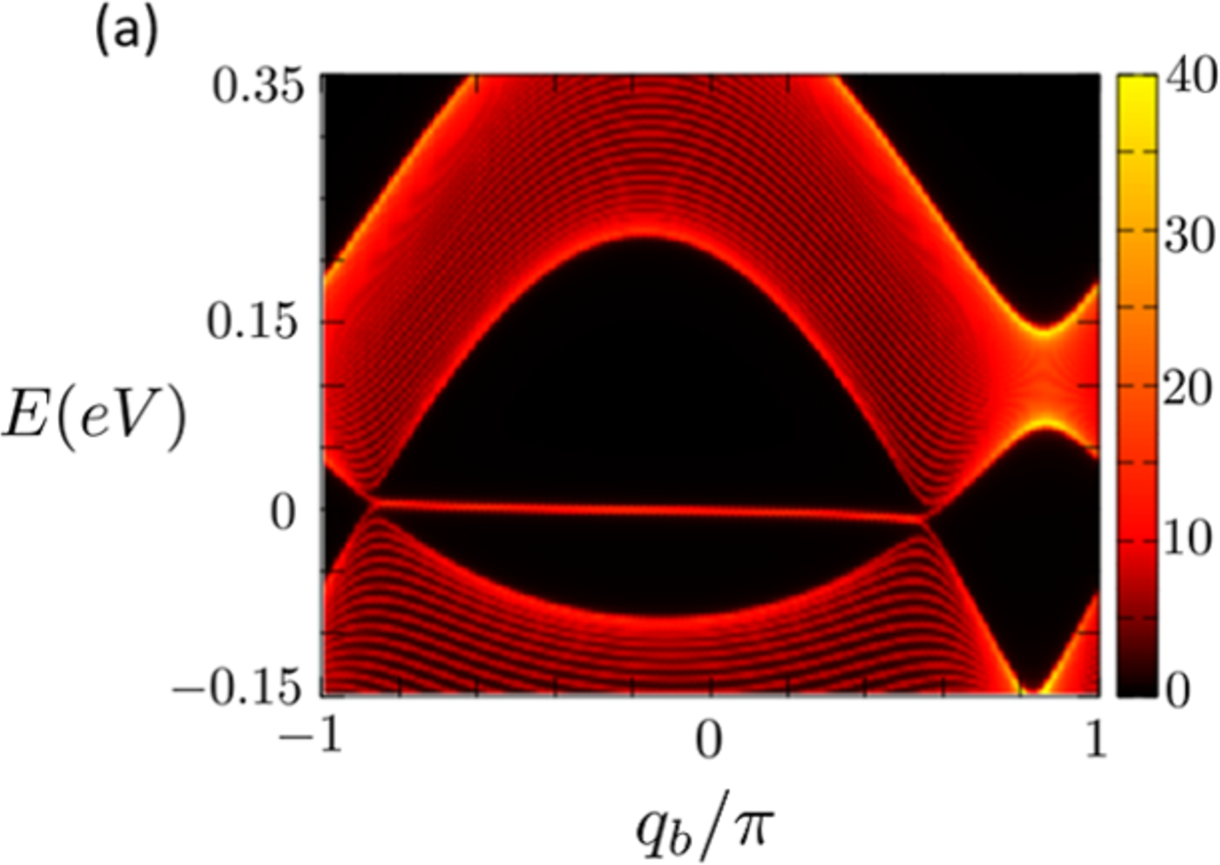}
\includegraphics[width=70mm]{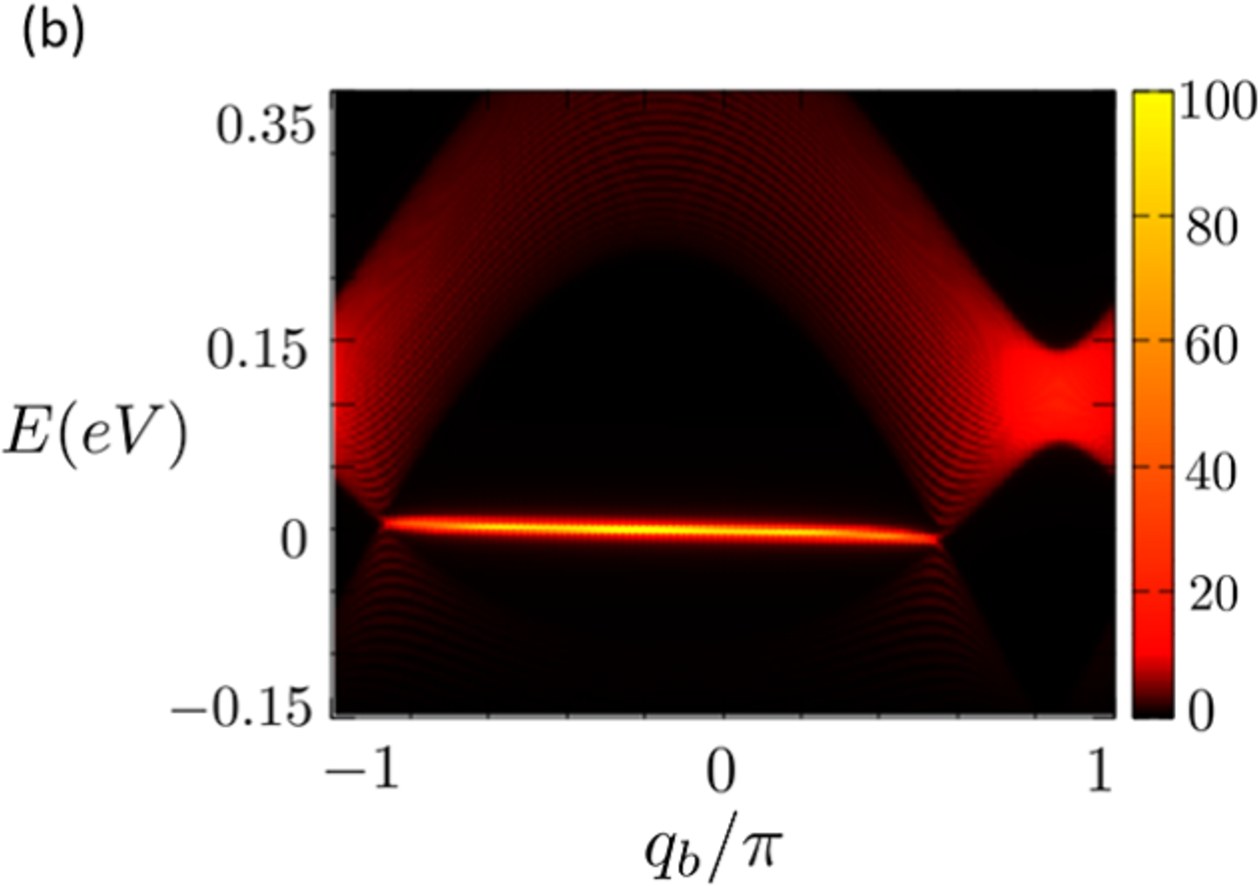}
\includegraphics[width=70mm]{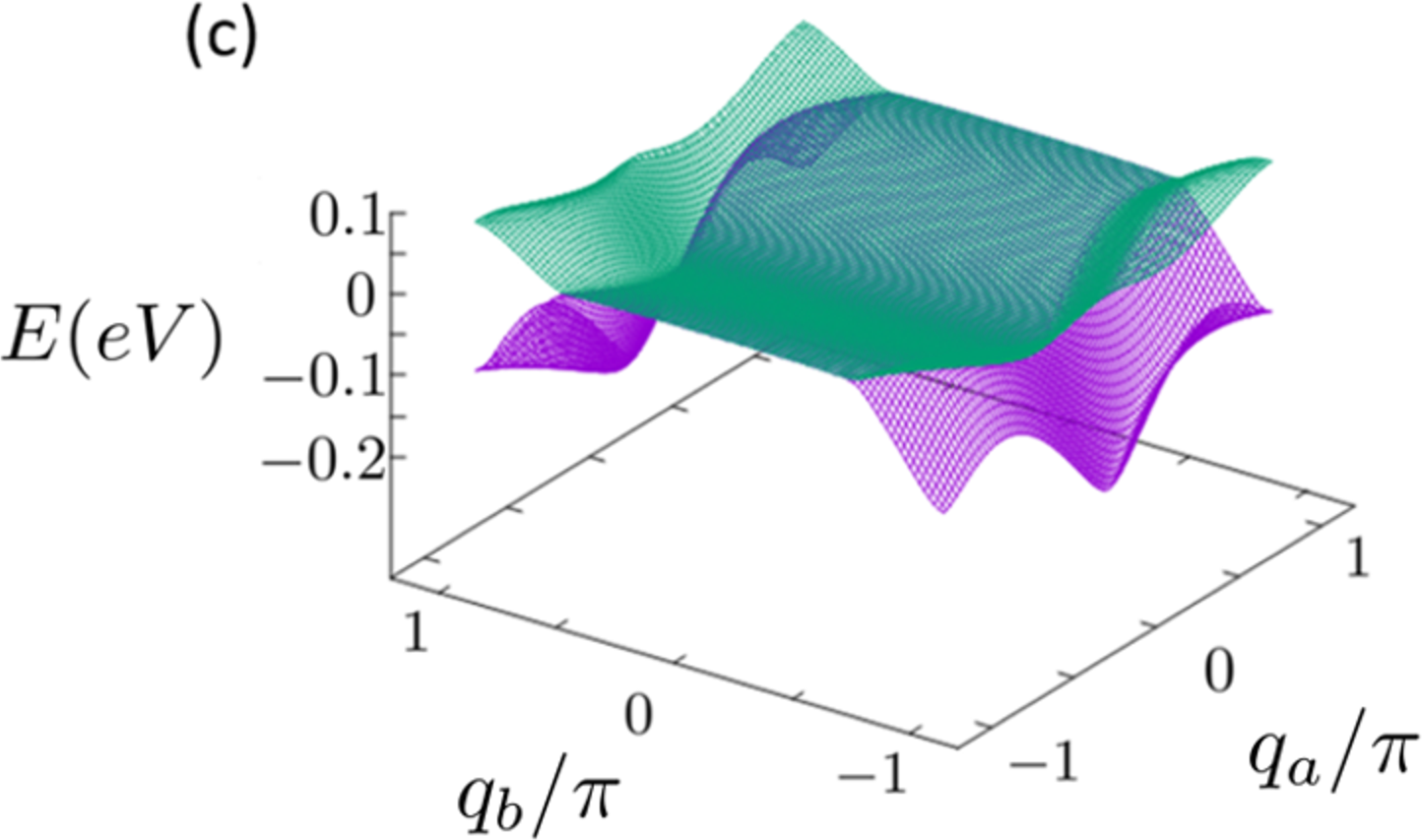}
\end{center}
\caption{(Color online)
$(a)$ Energy spectral weight at $q_{a}=-\pi/2$, where the horizontal axis is wavenumber $q_{b}$ and the vertical axis is the energy measured from the Fermi energy.
$(b)$ Local spectral weight of site 1 for the left edge. 
The edge states connect the two Dirac points directly. 
$(c)$ Energy dispersions (green upper and purple lower surfaces) that become the edge states between the two Dirac nodal lines in the $q_{a}$-$q_{b}$ plane.
}
\label{f1}
\end{figure}

Next, we investigate the LDOS $D_{i,\alpha}(E)$ at unit cell $i$ and site $\alpha$ for energy $E$. It is given by
\begin{equation}
D_{i,\alpha}(E)=\frac{1}{N_{a}N_{b}}\sum_{\textbf{q}} A_{i,\alpha}(\textbf{q},E)~,
\end{equation}
where $A_{i,\alpha}(\textbf{q},E)$ is the site-dependent local spectral weight [Eq. (7)].
The DOS is given by $D(E) = \frac{1}{N_{c}} \sum_{i,\alpha}D_{i,\alpha}(E)$.
Figure 8(a) shows the energy dependence of LDOS $D_{i,\alpha}(E)$ at the left edge in the energy range from $-0.15$ to $0.35$ eV. 
In this calculation, $N_{a}=200$, $N_{b}=200$, and $N_{c}=30$. 
LDOS at edges have logarithmic peaks near the Fermi energy, which is due to the quasi-1D energy dispersion of the edge states shown in Fig. 7. 
At the left edge, the edge states almost entirely consist of the sites 1 and 2 components.
The LDOSs of the sites 1 and 2 at the left edge are almost the same.
At the right edge, on the other hand, {we obtain the same $D_{i,\alpha}(E)$ as that at the left edge, but the green line shows the LDOS of site 3.
LDOS is $9.10$ [eV$^{-1}$] at the Fermi energy.
Figure 8(b) shows the energy dependence of the DOS. The DOS of the system in the presence of edges is shown by a purple line, and is calculated for $N_{a}=200$, $N_{b}=200$, and $N_{c}=30$. The DOS of the system in the absence of edges is shown by a green line and calculated for $N_{a}=100$, $N_{b}=300$, and $N_{c}=300$.
We obtain the contributions of both the quasi-1D edge states and the valley-like structure of the Dirac nodal line semimetal.

\begin{figure}[htpb]
\begin{center}
\includegraphics[width=70mm]{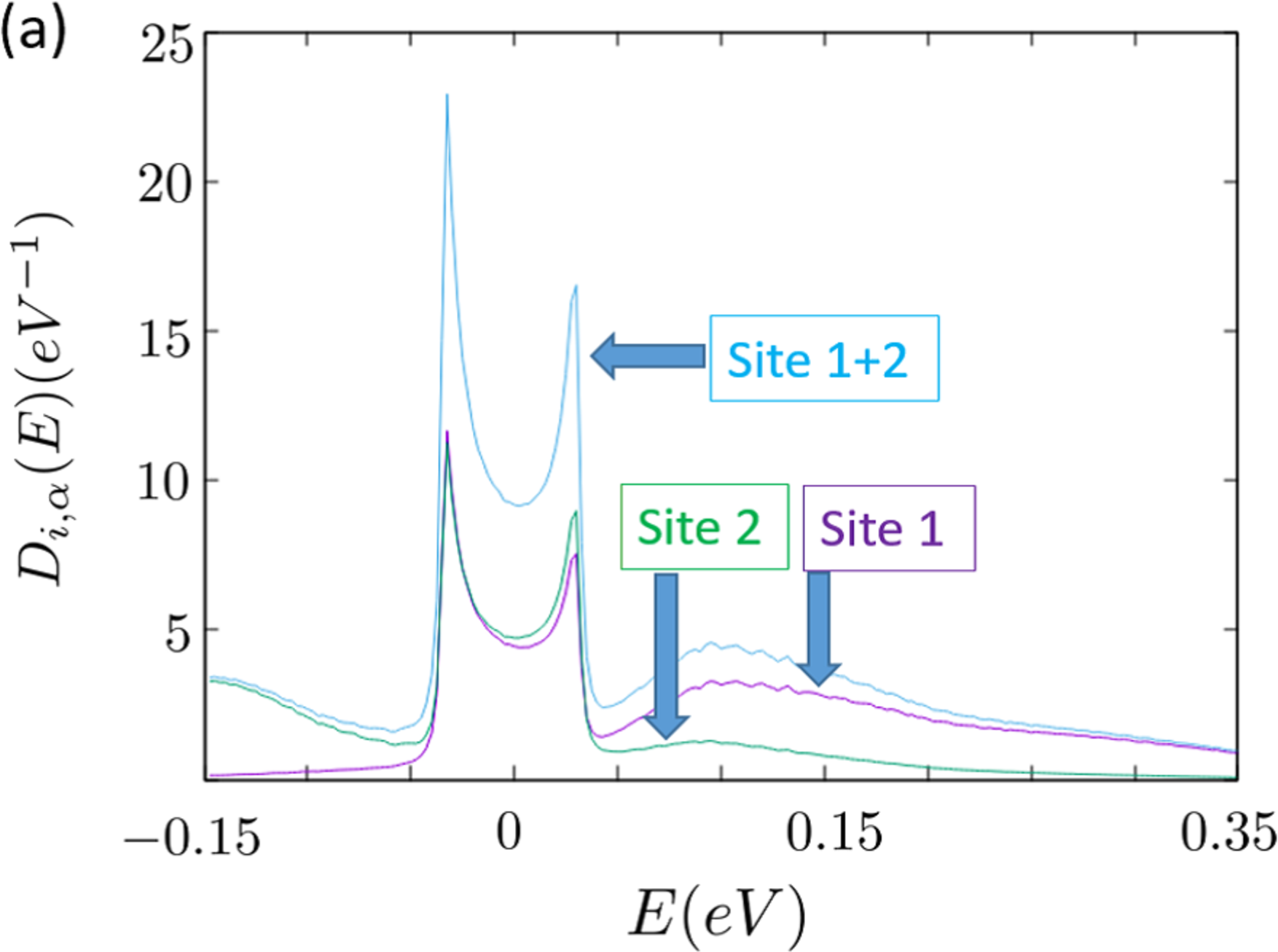}
\includegraphics[width=70mm]{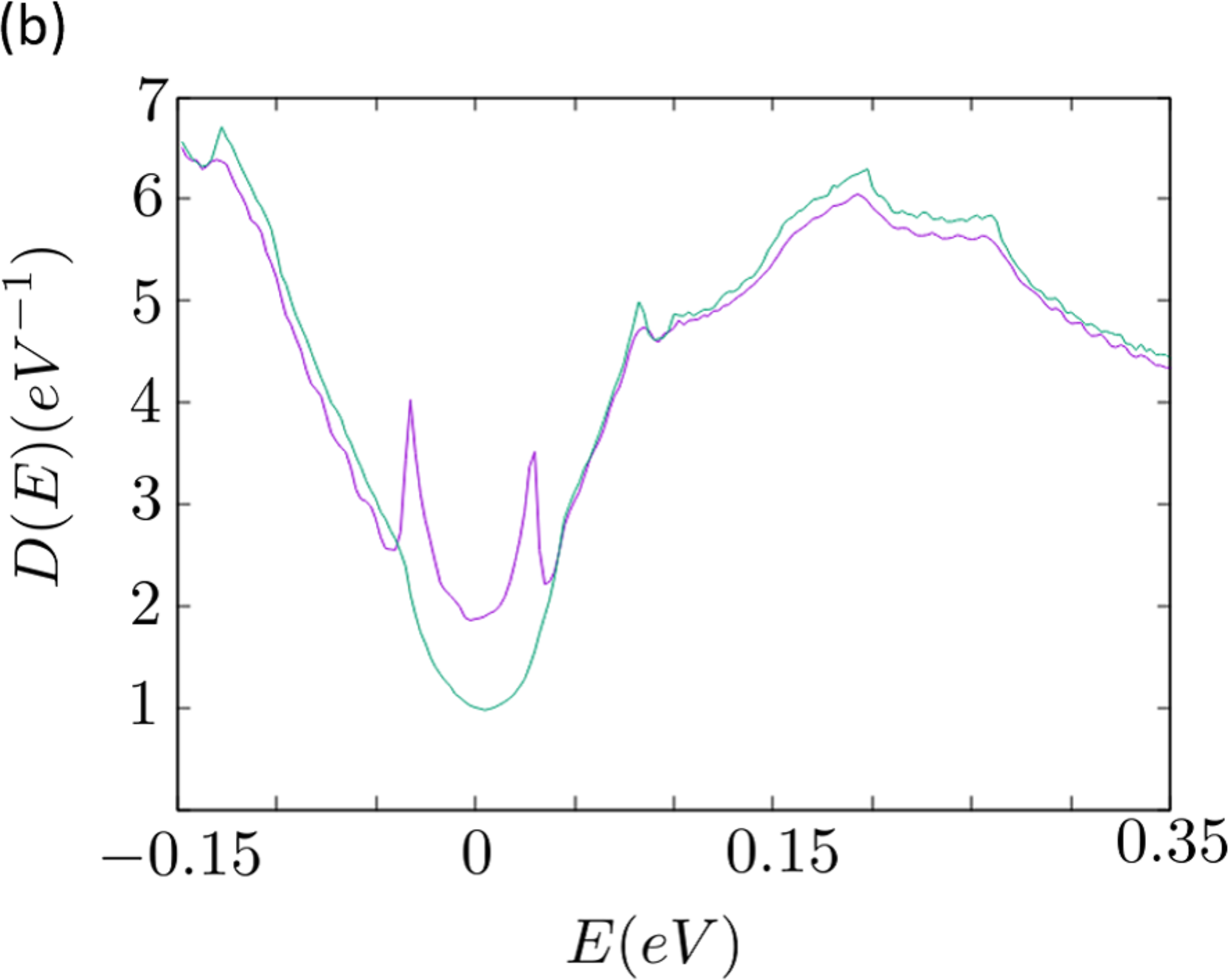}
\end{center}
\caption{(Color online)
(a) Energy dependence of the LDOS at the left edge. 
The LDOS of site 1 at the left edge is shown by a purple line. 
The LDOS of site 2 at the left edge is shown by a green line. 
Their sum is shown by a blue line. 
Logarithmic peaks that result from quais-1D energy dispersion exist near Fermi level. 
(b) Energy dependence of the DOS. 
The DOS of a cylindrical system with edges is shown by a purple line, where the logarithmic peaks appear near the Fermi energy. 
The DOS of the system without edges is shown by a green line. }
\label{f1}
\end{figure}

\section{Effects of Spin-Orbit Coupling}

Figure 9(a) shows the spectral weight at $q_{a}=-\pi/2$ for the SOC $\lambda=0.1$, where the horizontal axis is wavenumber $q_{b}$ and the vertical axis is the energy measured from the Fermi energy. 
In this calculation, $N_{a}=200$, $N_{b}=200$, and $N_{c}=40$. 
The SOC constant $\lambda=0.1$ incorporates a hopping energy of approximately $4$ meV. 
The bulk energy dispersion has the energy gap of approximately $20$ meV.
The edge states split at the two Dirac points and cross at the time reversal invariant momentum (TRIM).
Figure 9(b) shows schematic diagrams of the helical property of the edge states, which is confirmed by the analysis of the spin-dependent spectral weight.
Thus, we find that [Pt(dmdt)$_{2}$] with finite SOC is a topological material.
The Berry curvature and topological number are discussed in the next section.

\begin{figure}[htpb]
\begin{center}
\includegraphics[width=80mm]{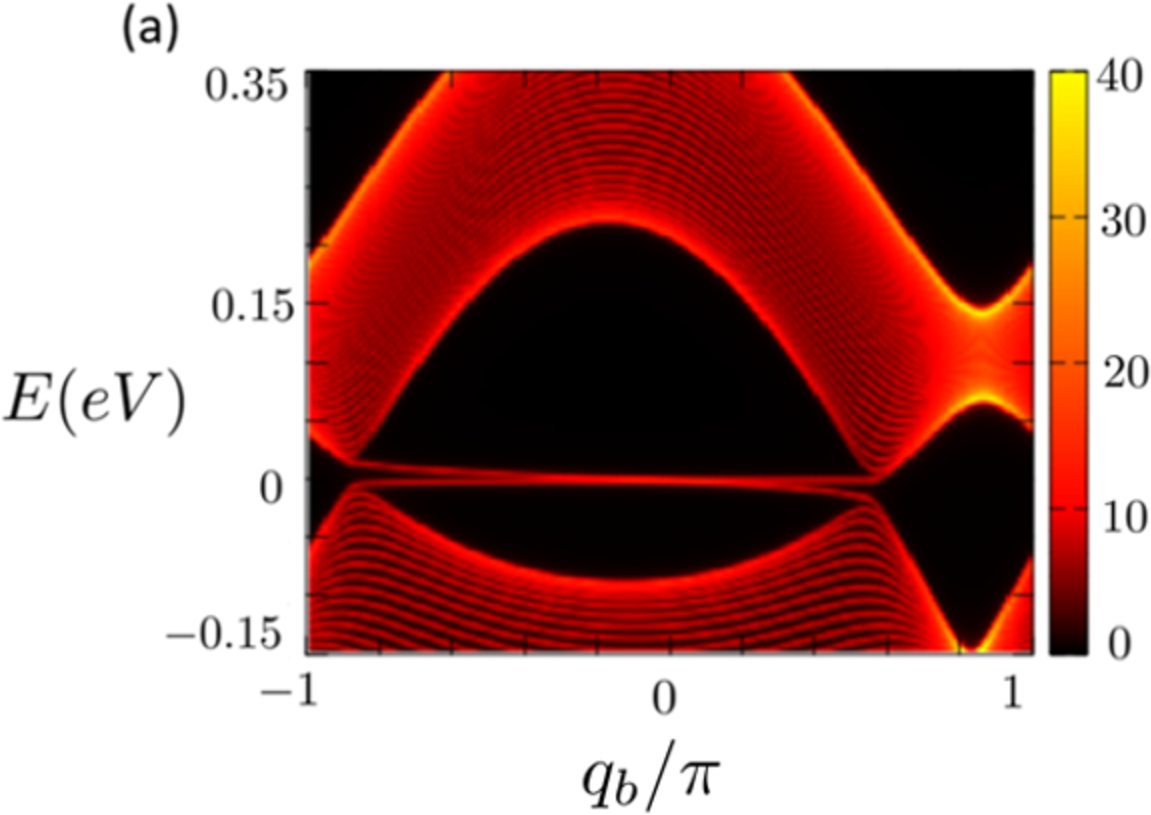}
\includegraphics[width=80mm]{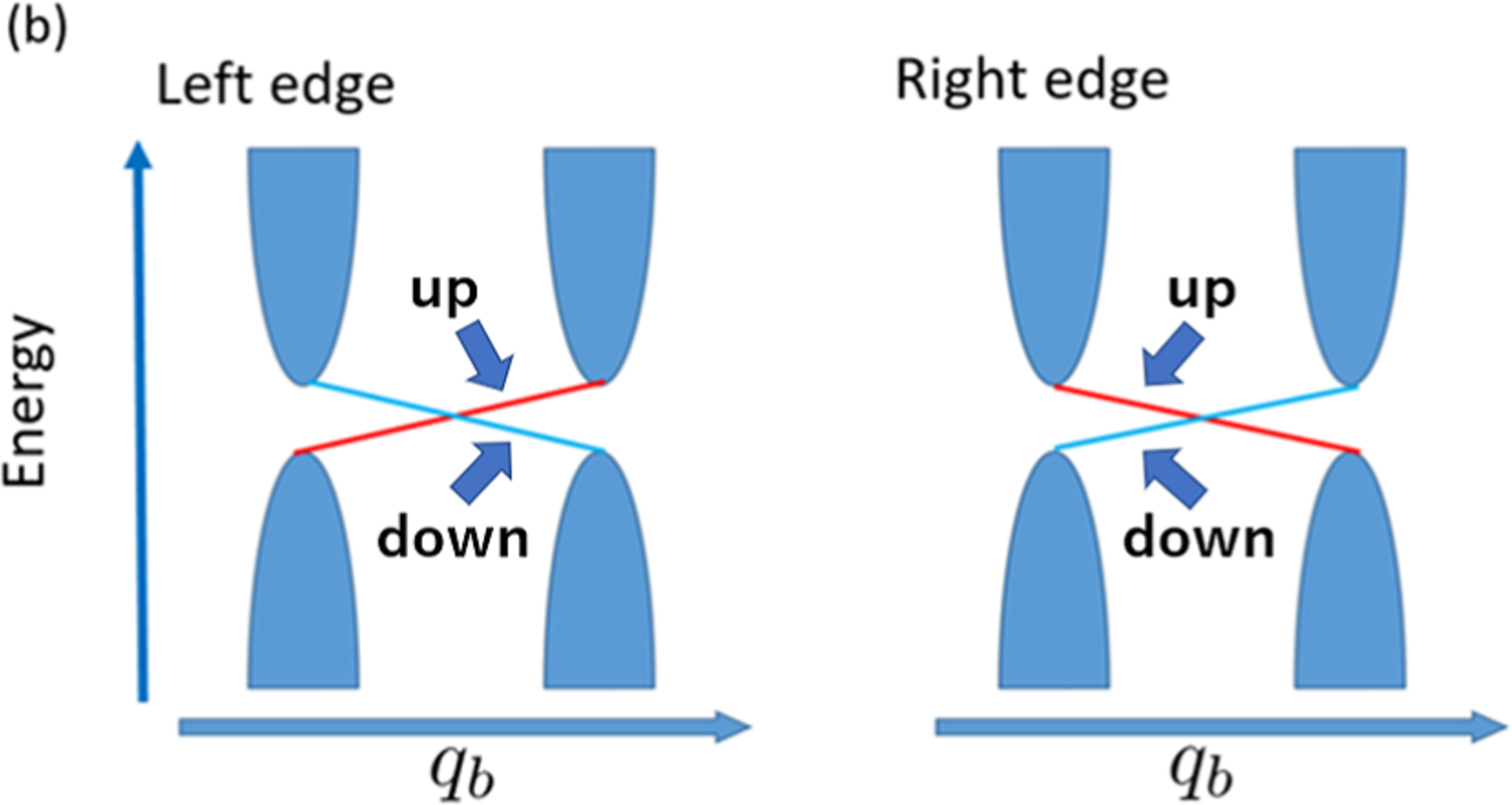}
\end{center}
\caption{(Color online)
(a) Spectral weight at $q_{a}=-\pi/2$.
The horizontal axis is wavenumber $q_{b}$ and the vertical axis is the energy measured from the Fermi energy.
(b) Schematic diagrams of the helical edge state, where the red (blue) line represents the edge state with up (down) spin.}
\label{f1}
\end{figure}

Figure 10 shows the energy dependence of the LDOS in the case of $\lambda=0.1$ at the edge. 
In this calculation, $N_{a}=200$, $N_{b}=200$, and $N_{c}=30$.
The logarithmic peaks near the Fermi energy are gradually suppressed as the SOC increases, since the quasi-1D dispersions of the helical edge state are bent.
The DOS at the Fermi energy, however, slightly increases with the SOC.

\begin{figure}[htpb]
\begin{center}
\includegraphics[width=60mm]{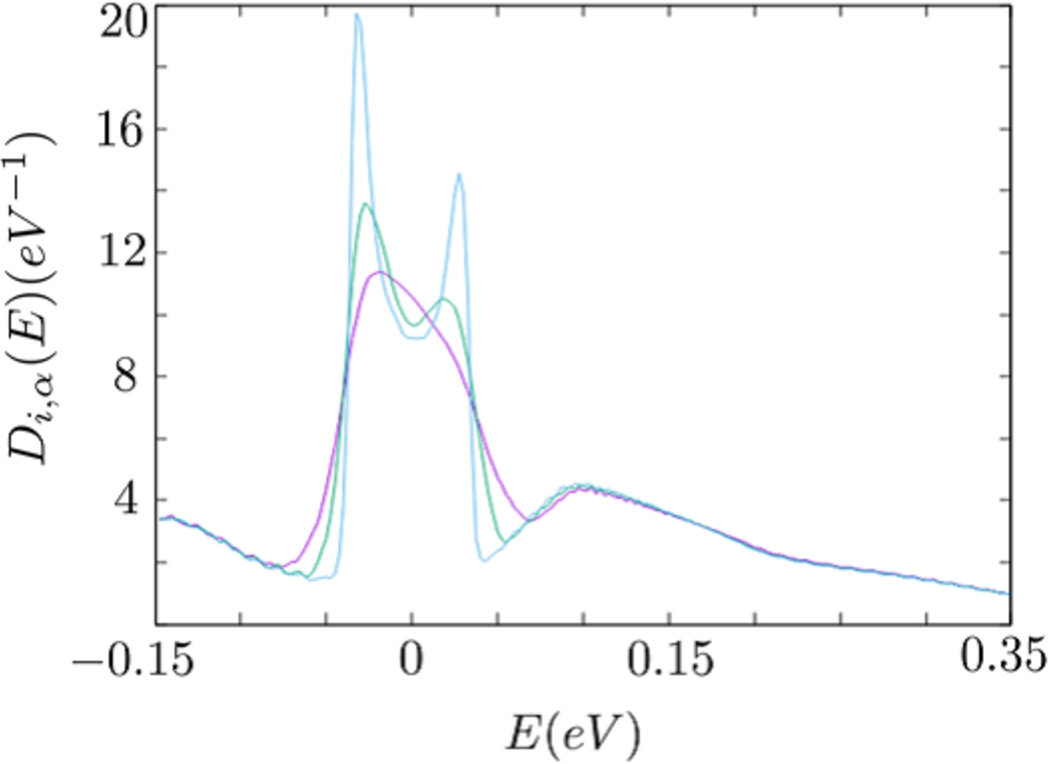}
\caption{(Color online)
Energy dependence of LDOS at the left edge for $\lambda=0.1$, $0.3$, and $0.5$. }
\end{center}
\label{f1}
\end{figure}

It is expected that [Pt(dmdt)$_{2}$] will become a topological insulator when the energy gaps induced by the SOC at the Dirac nodal lines are larger than the energy scale of the pockets.
Figure 11 shows the Fermi surfaces in the cases of $\lambda=0.0$, $0.1$, $0.2$, and $0.3$.
We find that the electron and hole pockets are separated and become smaller as $\lambda$ increases. 
When $\lambda \ge 0.39$, the electron and hole pockets vanish completely and [Pt(dmdt)$_{2}$] becomes a topological insulator.
This result indicates that the electron and hole pockets remain for a realistic SOC as shown by first-principles calculation\cite{Zhou2019} ($\lambda \sim 0.05$), although [Pt(dmdt)$_{2}$] is a topological material.

\begin{figure}[htpb]
\begin{center}
\includegraphics[width=80mm]{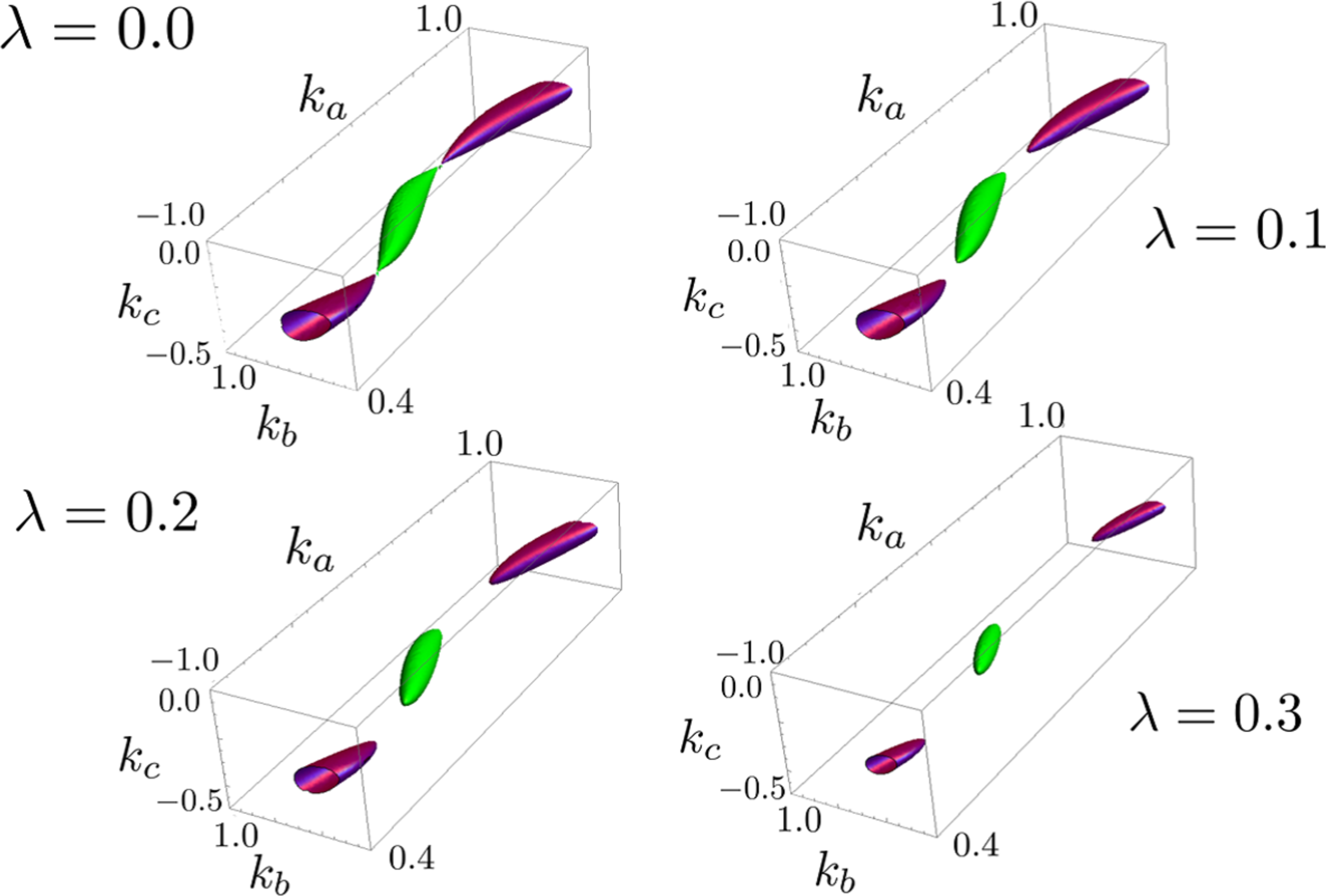}
\end{center}
\caption{(Color online)
Fermi surface for $\lambda=0.0$, $0.1$, $0.2$, and $0.3$.
The plot ranges are $-1 \le k_{a}/\pi \le 1$, $0.4 \le k_{b}/\pi \le 1$, and $-0.5 \le k_{c}/\pi \le 0$.
The green surfaces inside BZ and the purple ones on the boundary of BZ represent the electron and hole pockets, respectively.}
\label{f1}
\end{figure}

\section{Topological Properties}

We calculate the Berry curvature and the Chern number to determine whether [Pt(dmdt)$_{2}$] is a topological material.
The Berry curvature\cite{Berry} and the Chern number\cite{TKNN} are respectively defined by
\begin{equation}
{\small
\textbf{B}_{n,\sigma}(\textbf{k})=-i\sum_{m(\neq n)}\frac{\bra{\textbf{k},n,\sigma}\nabla_{\textbf{k}}H\ket{\textbf{k},m,\sigma}\times\bra{\textbf{k},m,\sigma}\nabla_{\textbf{k}}H\ket{\textbf{k},n,\sigma}}{(\epsilon_{\textbf{k},n,\sigma}-\epsilon_{\textbf{k},m,\sigma})^{2}}
} 
\end{equation}
and
\begin{equation}
Ch_{n,\sigma}=\frac{1}{2\pi}\int\int_{BZ}\textbf{B}_{n,\sigma}(\textbf{k})\cdot d\textbf{S}~,
\end{equation}
where $\textbf{k}$ is the wavenumber, $m$ and $n$ are band indices, and the $\sigma$ is spin index. 
$\ket{\textbf{k},m,\sigma}$ is the eigenvector of the Hamiltonian [Eqs. (2)-(4).] 
Figure 12 shows the $k_{b}$-$k_{c}$ dependence of the $a$-component of the Berry curvature of band 1 with up spin (a) and down spin (b) at $k_{a}=-\pi/2$. 
The Chern number satisfies $Ch_{1,\uparrow}=1$ and $Ch_{1,\downarrow}=-1$. 
Thus, the spin Chern number of [Pt(dmdt)$_{2}$] is nonzero. 

\begin{figure}[htpb]
\begin{center}
\includegraphics[width=60mm]{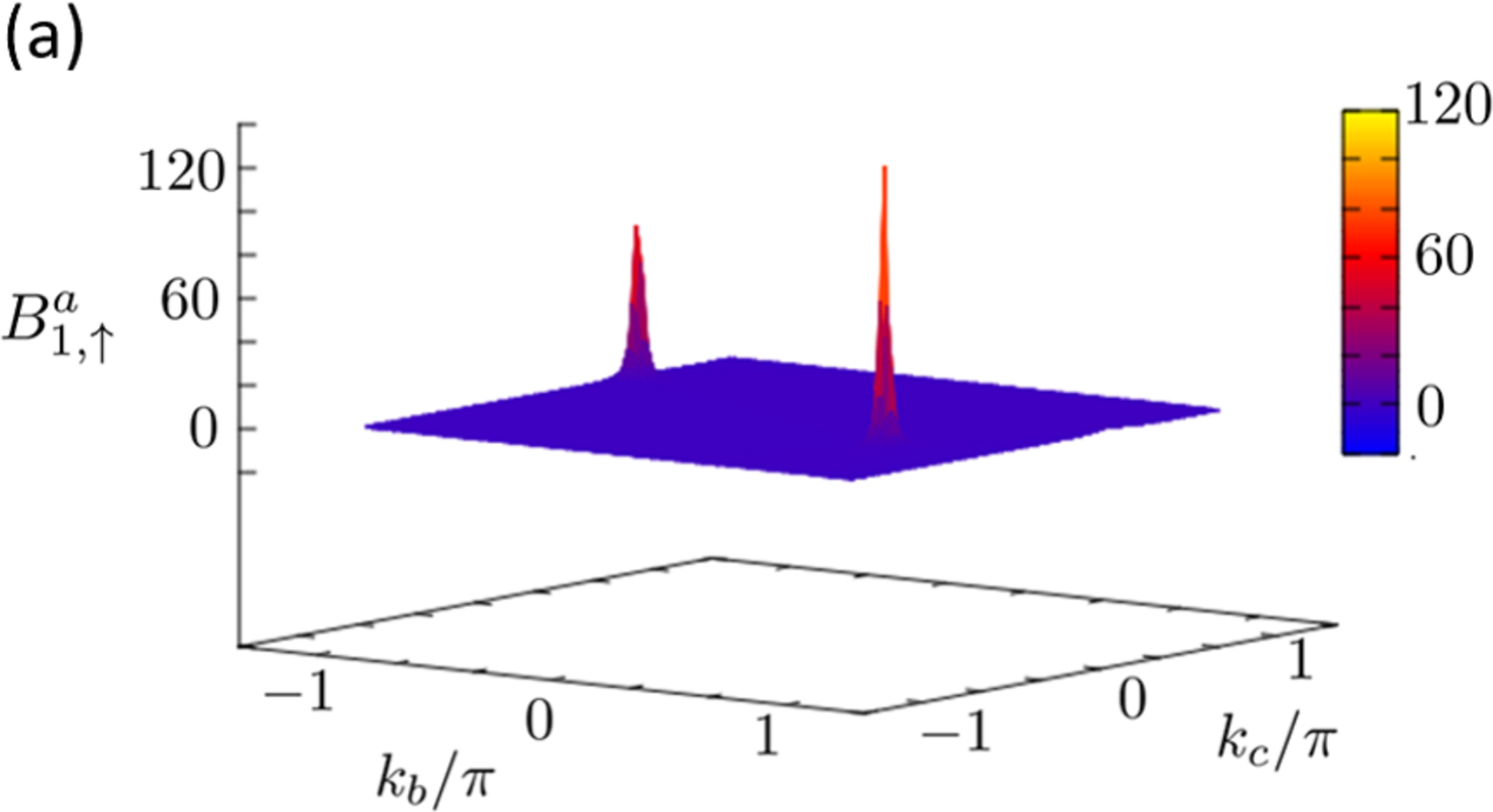}
\includegraphics[width=60mm]{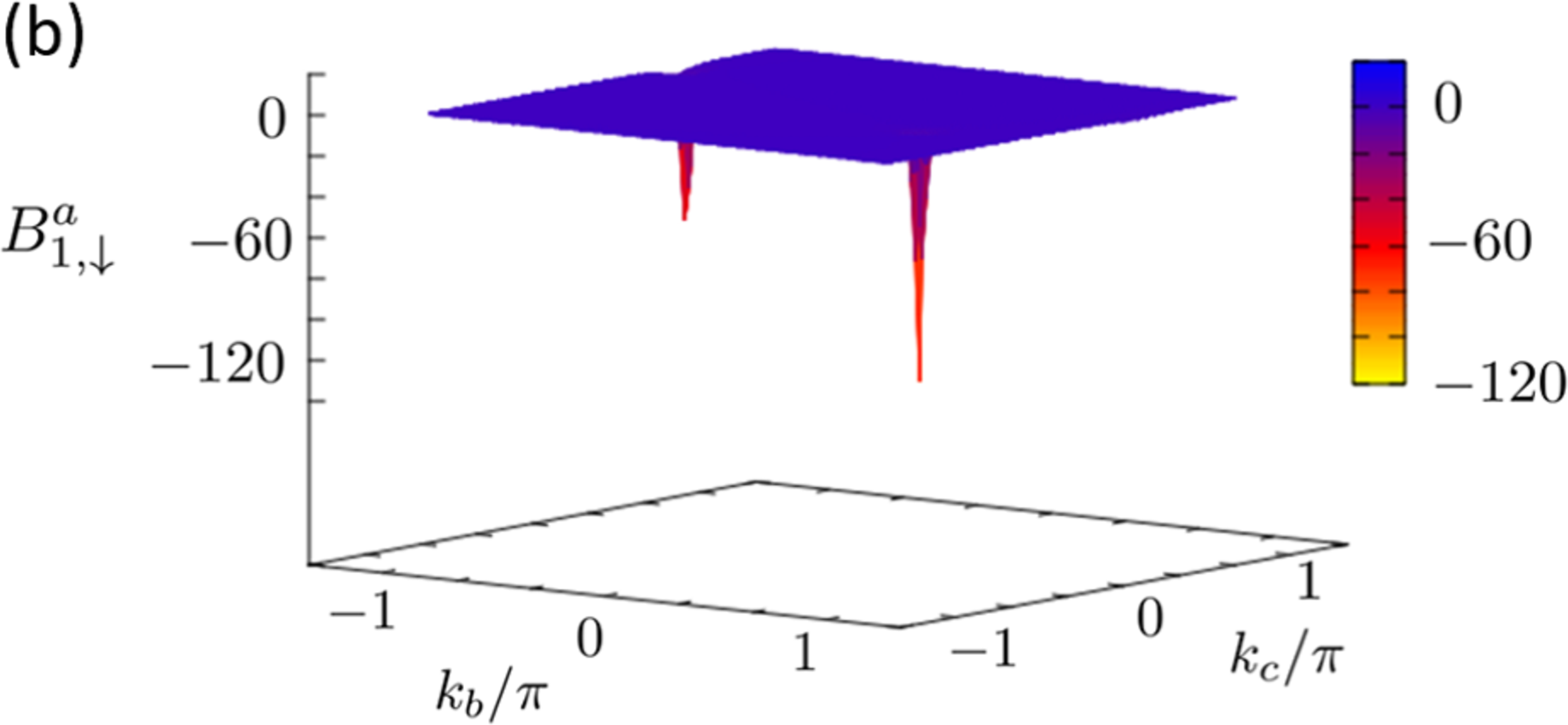}
\end{center}
\caption{(Color online)
$k_{b}$-$k_{c}$ dependence of $a$-component of the Berry curvature of band 1 at $k_{a}=-\pi/2$ in first BZ with up spin (a) and down spin (b).
The SOC constant is $\lambda =0.1$.
}
\label{f1}
\end{figure}

The topological number $\nu_{0}(\nu_{1},\nu_{2},\nu_{3})$ is defined as follows:\cite{FuKane,FuKaneMele,Moore,Roy,QiReview}
\begin{eqnarray}
(-1)^{\nu_{0}}\equiv \prod_{k_{a},k_{b},k_{c}=0,\pi}\delta_{k_{a},k_{b},k_{c}} \nonumber \\
(-1)^{\nu_{1}}\equiv \prod_{k_{b}=0,\pi}\prod_{k_{c}=0,\pi}\delta_{\pi,k_{b},k_{c}} \nonumber \\
(-1)^{\nu_{2}}\equiv \prod_{k_{a}=0,\pi}\prod_{k_{c}=0,\pi}\delta_{k_{a},\pi,k_{c}} \nonumber \\
(-1)^{\nu_{3}}\equiv \prod_{k_{a}=0,\pi}\prod_{k_{b}=0,\pi}\delta_{k_{a},k_{b},\pi}~,
\end{eqnarray}
where 
$\delta_{k_{a},k_{b},k_{c}}=\prod_{n=1}^{N_{filled}}\eta_{2n}(k_{a},k_{b},k_{c})$ with $\textbf{k}$ at the TRIM.
$\eta_{2n} =\pm 1$, which is called the parity eigenvalue, is obtained by acting the space inversion matrix on the eigenstate of band $2n$.
Because the Hamiltonian and the space inversion matrix are commutative at the TRIM, they can be diagonalized simultaneously there.
$\delta_{k_{a},k_{b},k_{c}}$ is calculated by taking the product of $\eta_{2n}$ for occupied bands. 
The space inversion matrix of the present three-orbital model [Eqs. (2) and (3)] describing the electronic states of [Pt(dmdt)$_{2}$] is given by
\begin{eqnarray}
P(\textbf{k})=
\left(
\begin{array}{ccc}
1 & 0 & 0 \\
0 & 0 & -e^{-i(k_{b}+k_{c})} \\
0 & -e^{i(k_{b}+k_{c})} & 0
\end{array}
\right) .
\end{eqnarray}
Since the Pt atom is the center of the space inversion, site 2 and 3 are exchanged by space inversion.
Figure 13 shows the $\delta_{k_{a},k_{b},k_{c}}$ of [Pt(dmdt)$_{2}$] at TRIMs. 
At $(k_{a},k_{b},k_{c})=(0,0,0),(\pi,0,0)$, $\delta_{k_{a},k_{b},k_{c}}$ is $1$. 
At other TRIMs, $\delta_{k_{a},k_{b},k_{c}}$ is $-1$. 
The bulk-edge correspondence is confirmed by the following quantities:\cite{FuKane,FuKaneMele,Moore,Roy,QiReview} 
\begin{eqnarray}
\xi_{a}(k_{b},k_{c})\equiv \prod_{k_{a}=0,\pi}\delta_{k_{a},k_{b},k_{c}} \\ \nonumber
\xi_{b}(k_{c},k_{a})\equiv \prod_{k_{b}=0,\pi}\delta_{k_{a},k_{b},k_{c}} \\ \nonumber
\xi_{c}(k_{a},k_{b})\equiv \prod_{k_{c}=0,\pi}\delta_{k_{a},k_{b},k_{c}}~.
\end{eqnarray}
These quantities are $\pm1$. 
The boundary that determines whether the edge state exists or not is found between TRIMs with different values. 

The edge state appears in the region where the TRIM gives the value $-1$. 
Figure 14 shows the values of $\xi_{a}(k_{b},k_{c})$, $\xi_{b}(k_{c},k_{a})$, and $\xi_{c}(k_{a},k_{b})$ at the corresponding TRIMs. 
The value $1$ is shown by blue filled circles and the value $-1$ is shown by red circles. 
The orange zones are the regions where the edge state appears. 
It is shown that the edge state does not appear at the edge perpendicular to the $a$-axis.
The edge state appears in the region along to the $k_{a}$ axis.
The case of Fig. 14(c) corresponds to the results in Sects. 4 and 5.

\begin{figure}[htpb]
\begin{center}
\includegraphics[width=50mm]{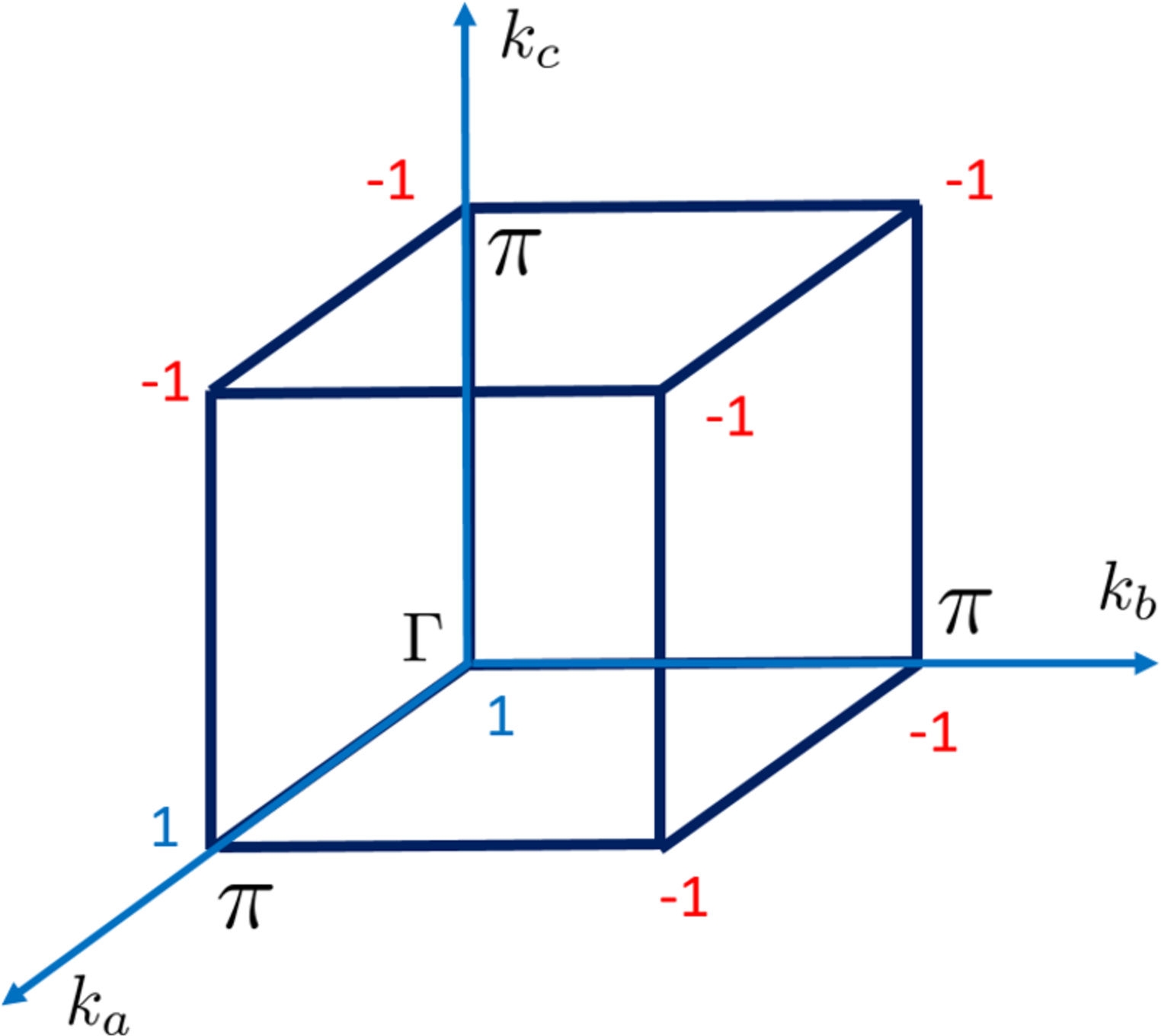}
\end{center}
\caption{(Color online)
$\delta_{k_{a},k_{b},k_{c}}$ at TRIMs, which is 1 at $(k_{a},k_{b},k_{c})$=$(0,0,0)$, $(\pi,0,0)$, and -1 at the other TRIMs.}
\label{f1}
\end{figure}

\begin{figure}[htpb]
\begin{center}
\includegraphics[width=30mm]{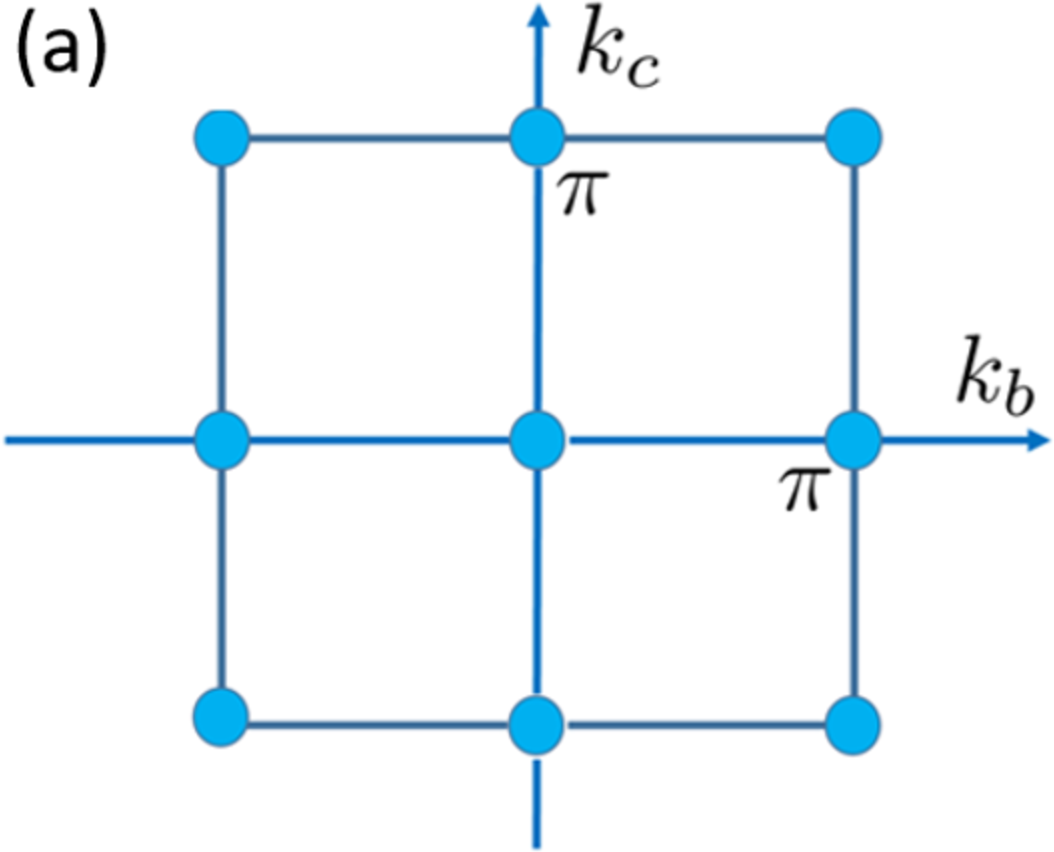}
\includegraphics[width=30mm]{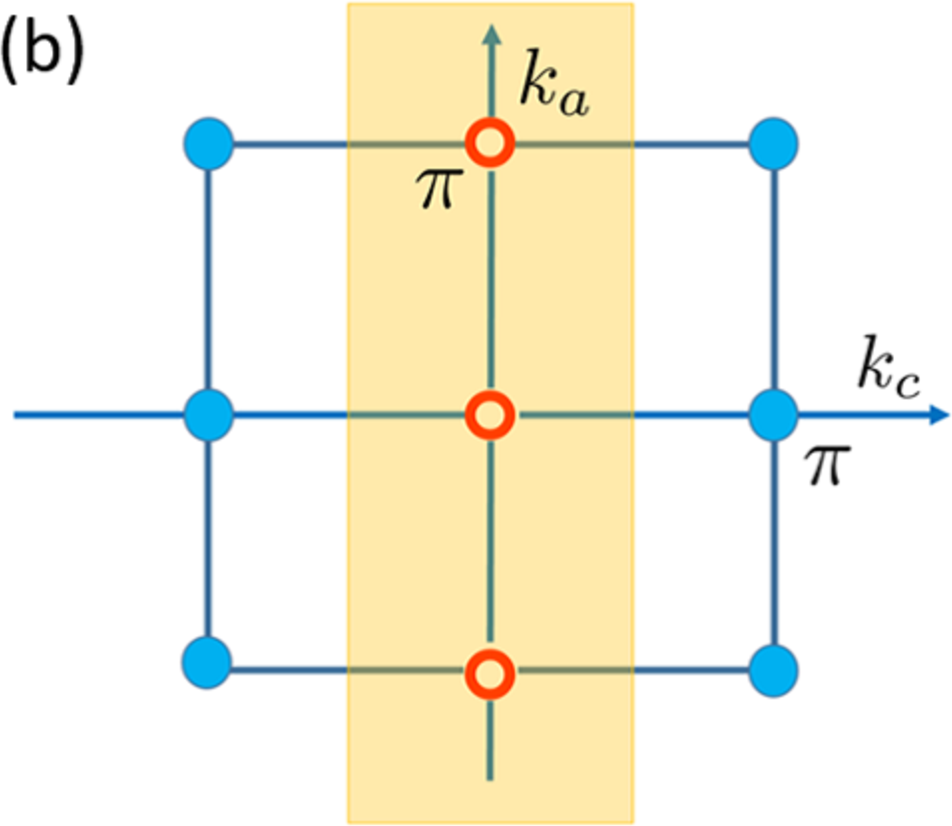}
\includegraphics[width=30mm]{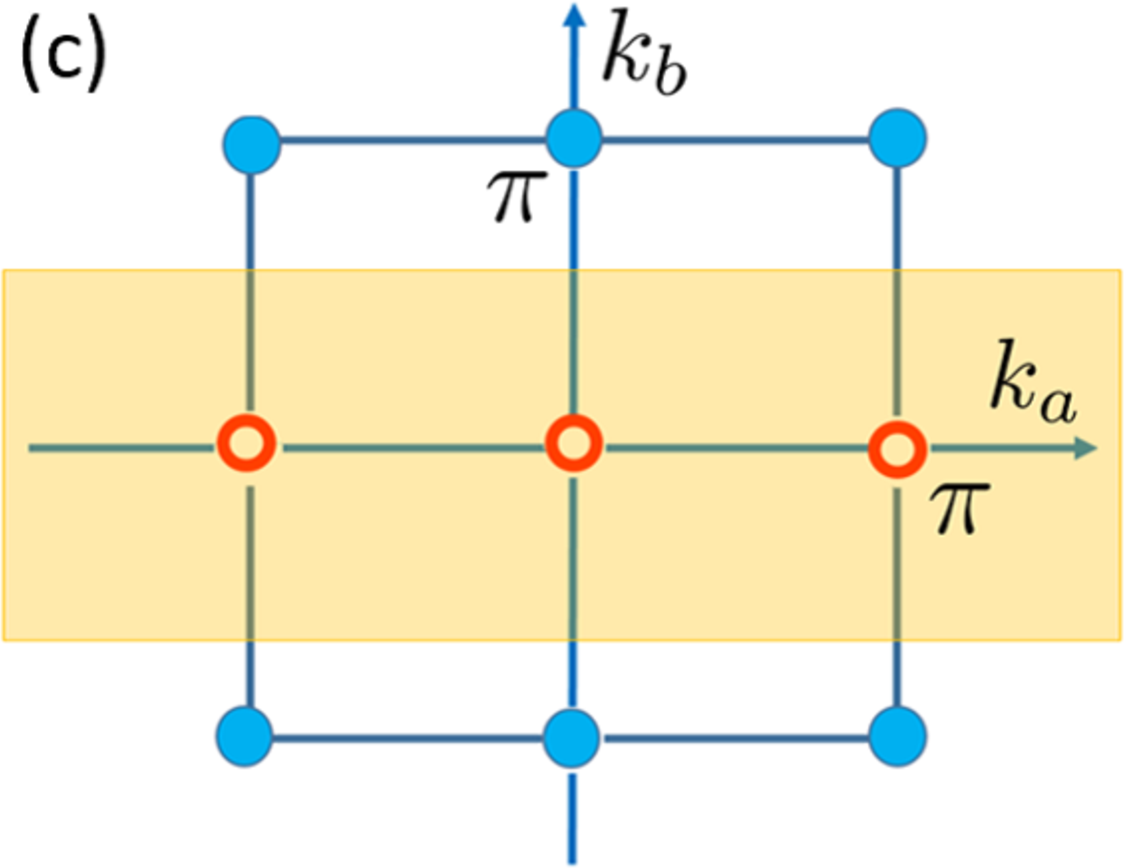}
\end{center}
\caption{(Color online)
The values of $\xi_{a}(k_{b},k_{c})$, $\xi_{b}(k_{c},k_{a})$ and $\xi_{c}(k_{a},k_{b})$ at corresponding TRIMs. The value $1$ is shown by blue filled circles and the value $-1$ is shown by red circles. 
The orange zones are the regions where the edge state appears.
}
\label{f1}
\end{figure}

Thus, [Pt(dmdt)$_{2}$] is a weak topological material characterized by topological number $0(100)$ with the pseudo-1D edge states.

\section{Magnetic Susceptibility}

The temperature dependence of magnetic susceptibility is given by\cite{Katayama2009} 
\begin{equation}
\chi(T) =\int_{-\infty}^{\infty}D(\epsilon)\left(-\frac{df}{d\epsilon}\right)d\epsilon~.
\end{equation}
Figure 15 shows the temperature dependence of magnetic susceptibility. 
When $\lambda < 0.39$, the magnetic susceptibility at a high temperature linearly decreases as the temperature decreases.
The magnetic susceptibility at a low temperature is almost constant and remains finite at $T=0$, which is due to the electron and hole pockets. 
For $\lambda=0.4$ (corresponding to approximately $0.017$ eV), the magnetic susceptibility linearly decreases to zero as the temperature decreases, because the electron and the hole pockets vanish for a large SOC as shown in Fig. 11. 
For $\lambda=0.8$ (corresponding to approximately $0.035$ eV), the magnetic susceptibility becomes zero below 100 K. 

The $T$-linear behaviors of the magnetic susceptibility for all $\lambda$ above 100 K shown in Fig. 15 are consistent with the experimental result at a high temperature, \cite{Zhou2019} supporting the existence of Dirac nodal line.
Below 100 K, the depletion of the magnetic susceptibility for $\lambda=0.8$ may appear consistent with the experimental result at a low temperature\cite{Zhou2019}. 
However, first-principles calculation predicts much smaller SOC ($\lambda \sim 0.05$, corresponding to approximately $0.002$ eV)\cite{Zhou2019}. 
Thus, the low-temperature behavior of the observed magnetic susceptibility cannot be explained only by the effect of the SOC.

\begin{figure}[htpb]
\begin{center}
\includegraphics[width=60mm]{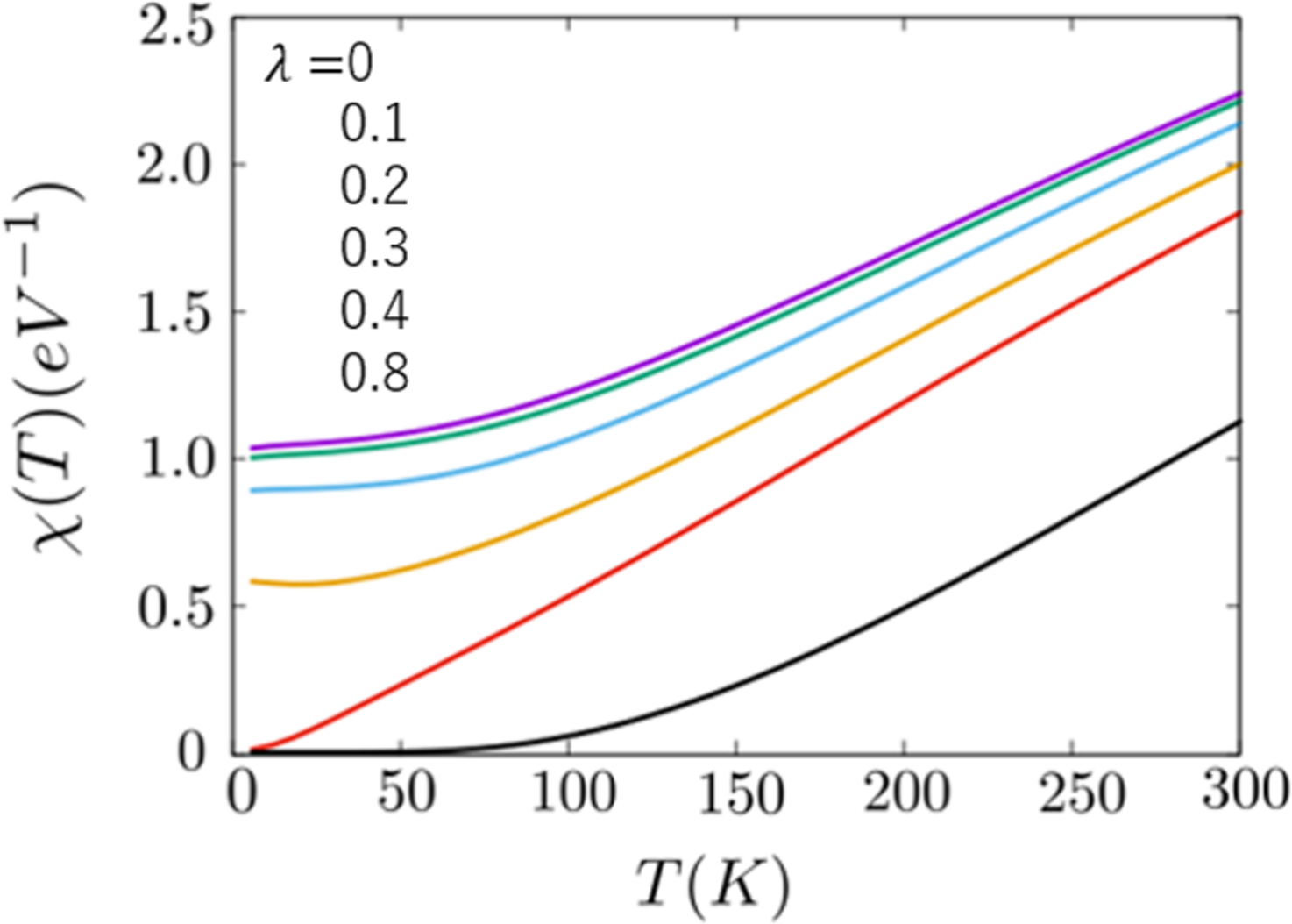}
\end{center}
\caption{(Color online)
Temperature dependence of magnetic susceptibility for $\lambda=0, 0.1, 0.2, 0.3, 0.4$, and $0.8$ (the purple, green, blue, yellow, red, and black lines, respectively).}
\label{f1}
\end{figure}

\section{Summary and Discussion}

We have constructed three-orbital tight-binding model based on the Wannier fitting of the first-principles calculation of [Pt(dmdt)$_{2}$].
The present model reproduces the three isolated bands near the Fermi energy obtained by first-principles calculation, where those bands include the Dirac nodal line and the Dirac nodal ring. 
We have found flat edge states between the Dirac nodal lines, which induce the characteristic logarithmic peaks of the LDOS near the Fermi energy.
We have calculated the magnetic susceptibility using the DOS in our model.
We have found that the $T$-linear behavior of the magnetic susceptibility at a high temperature is consistent with the experimental result at a high temperature\cite{Zhou2019} supporting the existence of Dirac nodal line, although the depletion of the magnetic susceptibility at a low temperature cannot be explained only by the effect of a realistic SOC.
We have shown that [Pt(dmdt)$_{2}$] is a topological nodal line semimetal with isolated electron and hole pockets in the presence of a realistic SOC.
If the SOC is larger than a threshold, those pockets vanish and the system becomes a weak topological insulator.

The present results for the SOC may be the key to finding weak topological insulators in the family of [Pt(dmdt)$_{2}$].
It is expected that the logarithmic peaks of the LDOS will give rise to edge magnetism at the $(0,0,1)$ surface in the presence of the short-range Coulomb interaction. 
It is also expected that the long-range Coulomb interaction, on the other hand, will affect the temperature dependences of the magnetic susceptibility and nuclear magnetic relaxation rate, as shown for 
$\alpha$-(BEDT-TTF)$_2$I$_3$\cite{Hirata2016,Hirata2017}.
The details of the calculation will be provided elsewhere.
Furthermore, the pseudo-1D helical edge states of [Pt(dmdt)$_{2}$] found in the present study have the potential to carry a highly directional spin current\cite{Noguchi}, since a spin Hall current exists in the presence of Fermi pockets\cite{Fuseya2015}.

\begin{acknowledgment}
The authors would like to thank S. Ishibashi, K. Kanoda, and H. Fukuyama for fruitful discussion. 
This work was supported by MEXT (JP) JSPJ (Grants No. 15K05166, 19H01846, 19J20677, and 17K05846).
The computation in this work has been done using the facilities of the Supercomputer Center, the Institute for Solid State Physics, the University of Tokyo.
\end{acknowledgment}

\appendix
\section{Simplified Models and Stability of Dirac Point}
We consider a two-dimensional three-orbital tight-binding model that includes the hopping energies $t_{1}$, $t_{2}$, and $t_{3}$, which are more than three times as large as the others, 
\begin{equation}
H_{\rm 2D}(\textbf{k})=
\left(
\begin{array}{ccc}
0 & -t_{1}e^{ik_{c}} & t_{1}e^{-ik_{b}} \\
-t_{1}e^{-ik_{c}} & 0 & t_{2}+t_{3}e^{-ik_{b}} \\
t_{1}e^{ik_{b}} & t_{2}+t_{3}e^{ik_{b}} & 0
\end{array}
\right)
\end{equation}
with $t_{1}=0.212$, $t_{2}=0.179$, and $t_{3}=0.201$ eV, where the spin-orbit interaction and the site potentials are ignored for simplicity.
The energy dispersion is shown in Fig. A$\cdot$1.
This model approximately reproduces the linear energy dispersion on the $k_{a}=-\pi/2$ plane shown in Fig. 4(a).

\begin{figure}[htpb]
\begin{center}
\includegraphics[width=80mm]{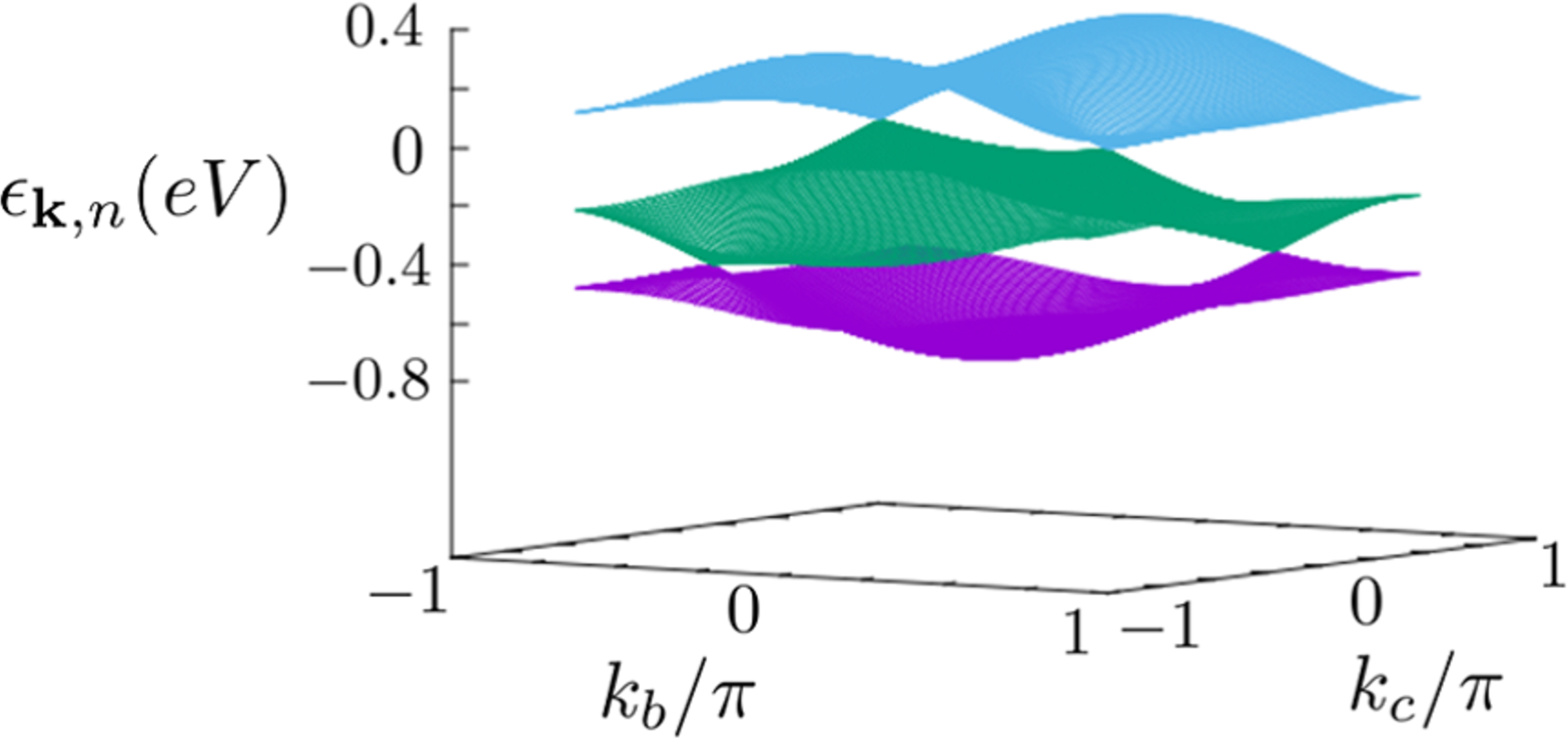}
\end{center}
\caption{(Color online)
Energy dispersion of a two-dimensional three-orbital tight-binding model $H_{\rm 2D}(\textbf{k})$.
Dirac cones exist between bands 1 and 2, and between the bands 2 and 3.}
\label{f1}
\end{figure}

This result indicates that the essential feature of the electron system in [Pt(dmdt)$_{2}$] is determined by $t_{1}$, $t_{2}$, and $t_{3}$. 
Thus, we investigate the stability of the gapless Dirac points based on $H_{\rm 2D}(\textbf{k})$. 
Since space inversion symmetry plays an important role in the stability of many Dirac electron systems, we consider the most simplified two-dimensional three-orbital tight-binding model given by 
\begin{equation}
{\rm H_S}(\textbf{k})=
\left(
\begin{array}{ccc}
0 & -t'e^{ik_{c}} & te^{-ik_{b}} \\
-t'e^{-ik_{c}} & 0 & t+te^{-ik_{b}} \\
te^{ik_{b}} & t+te^{ik_{b}} & 0
\end{array}
\right)~.
\end{equation}
When $t=t'$, this model has space inversion symmetry.
The space inversion symmetry is broken when $t\neq t'$.
We calculate the energy gap at the Dirac points between bands 1 and 2.
Figure A$\cdot$2(a) shows the $t'/t$ dependence of the energy gap, where we set $t=1$.
It is shown that the energy gap vanishes at $t'/t=1$, which means that the gapless Dirac point is protected by the space inversion symmetry. 
The energy dispersions near the Dirac cone for $t'/t=1$ and $t'/t=0.9$ are shown in Figs. A$\cdot$2(b) and A$\cdot$2(c), respectively.
Figure A$\cdot$2(b) shows the gapless Dirac cone, while Fig. A$\cdot$2(c) shows the Dirac point with a finite gap due to the inversion symmetry breaking with $t\neq t'$. 
\begin{figure}[htpb]
\begin{center}
\includegraphics[width=50mm]{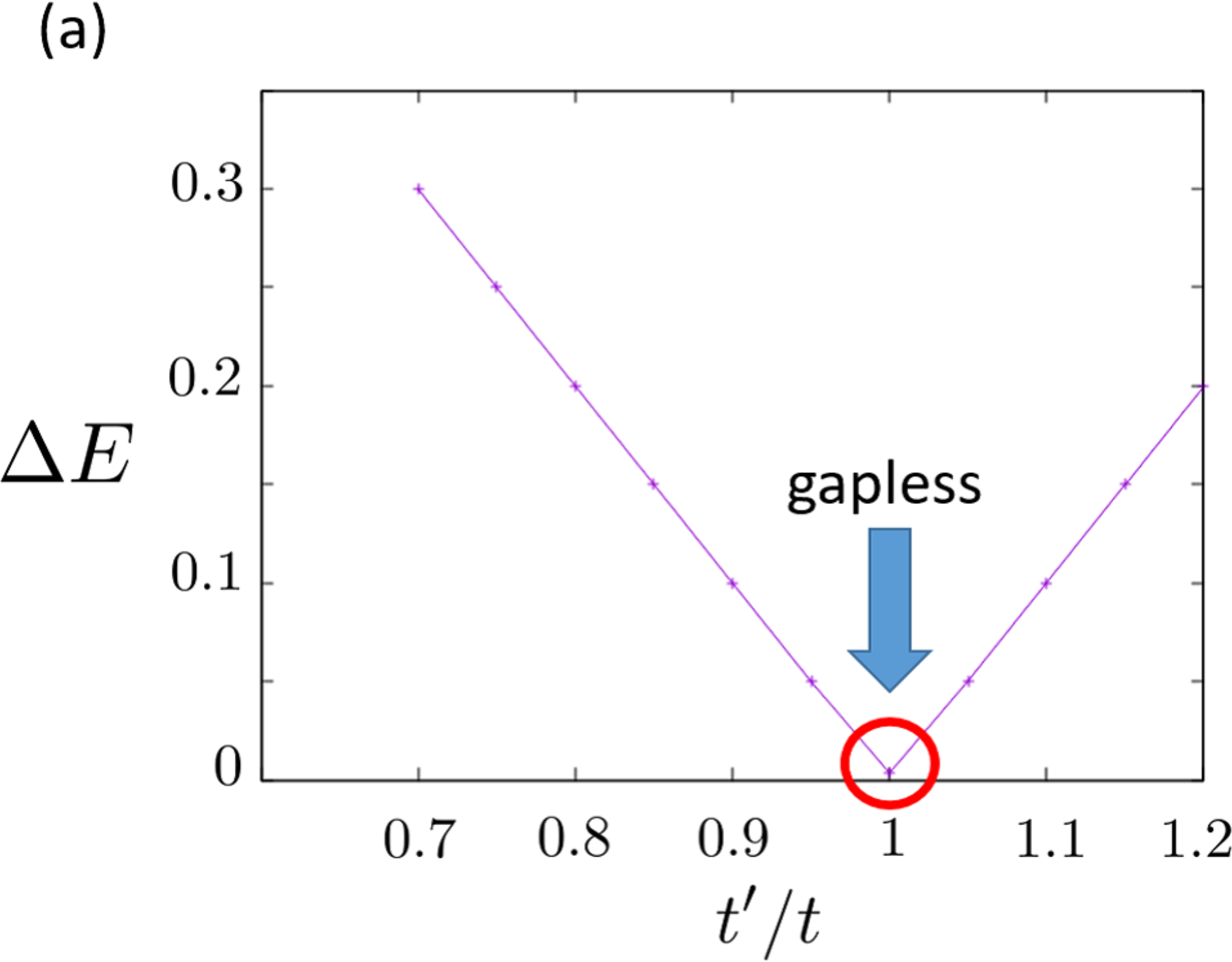}
\includegraphics[width=50mm]{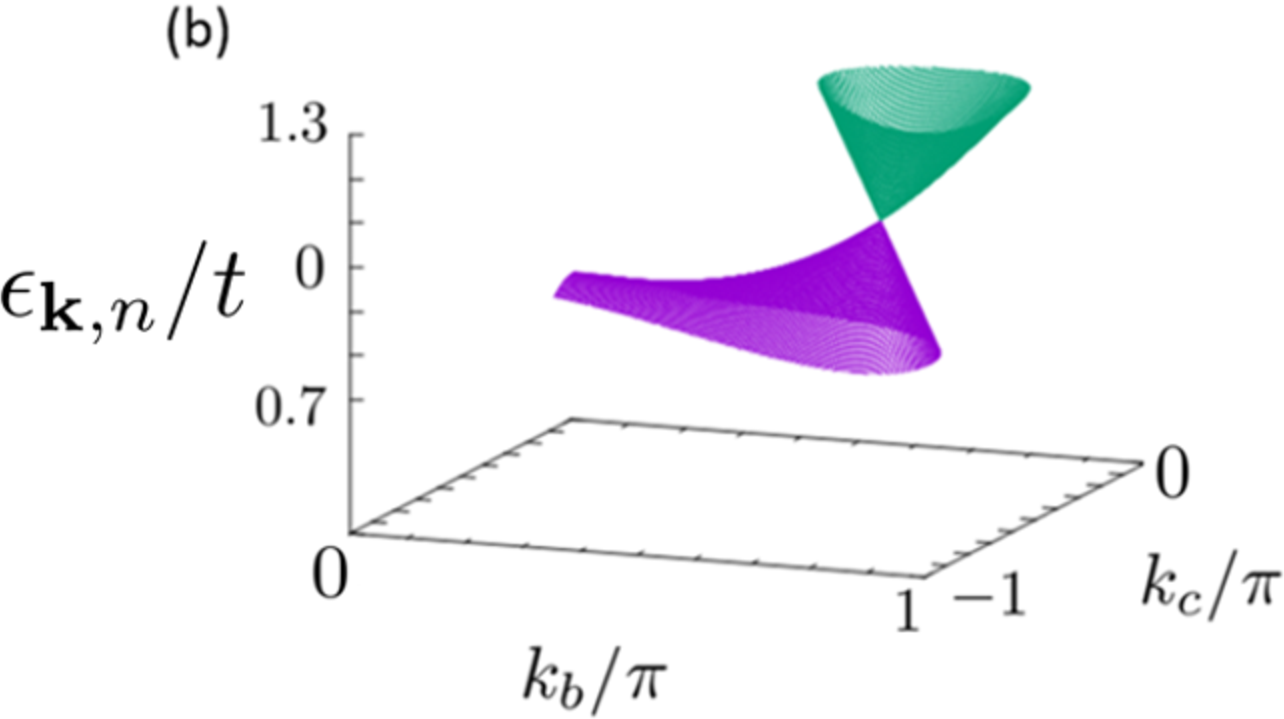}
\includegraphics[width=50mm]{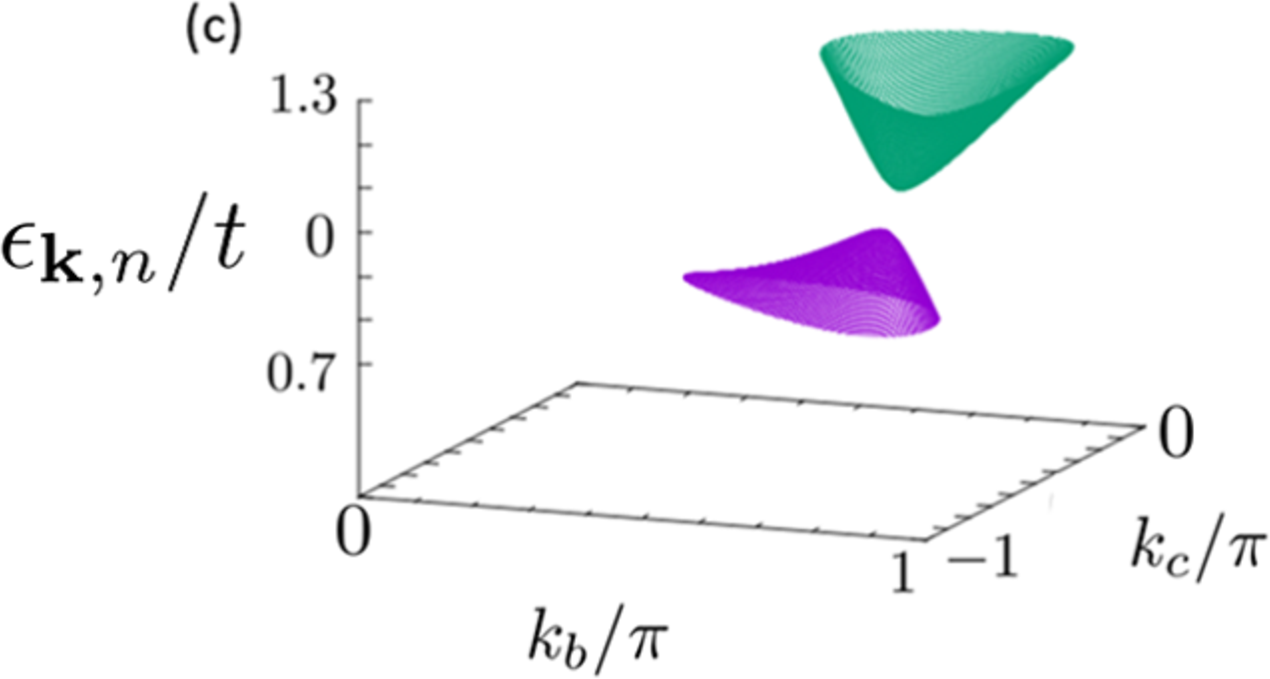}
\end{center}
\caption{(Color online)
(a)$t'/t$ dependence of the energy gap at the Dirac points in the most simplified two-dimensional three-orbital tight-binding model $H_{\rm S}$.
(b) The energy dispersion for ratio $t'/t=1$. 
Energy range is restricted near the Dirac cone for visibility. 
Wavenumber $k_{b}$ range is from 0 to 1 and $k_{c}$ range is from -1 to 0. 
(c) The energy dispersion for ratio $t'/t=0.9$. }
\label{f1}
\end{figure}
In the case with $t'/t=1$, the wavenumbers $(k_{b}, k_{c})$ of the Dirac points are determined by the following equation with $t=1$:
\begin{equation}
E^{3}+(2+4\cos{k_{b}})E+(2\cos{k_{c}}+2\cos{(k_{b}+k_{c})})=0~.
\end{equation} 
Therefore, the Dirac points satisfy following discriminant:
\begin{equation}
8(\cos{k_{b}}+2)^{3}+27(\cos{k_{c}}+\cos{(k_{b}+k_{c})})^{2}=0~.
\end{equation}

\appendix
\section{Correspondence to the Two-Orbital Model}
We show the relationship between the present three-orbital model [Eqs. (2) and (3)]
and the two-orbital model based on the extended H\"{u}ckel method\cite{Kato2020}. 
We consider the one-molecule Hamiltonian of Pt(dmdt)$_{2}$ including the hopping energies inside a Pt(dmdt)$_{2}$ molecule and the site potentials shown in Fig. B$\cdot$1. 

The one-molecule Hamiltonian is given by 
\begin{eqnarray}
H=
\left(
\begin{array}{ccc}
\Delta & -t_{1} & t_{1} \\
-t_{1} & 0 & t_{6} \\
t_{1} & t_{6} & 0
\end{array}
\right)
\end{eqnarray}
with $t_{1}=0.212$ and $t_{6}=0.042$. The potential energy of site 1 is $\Delta=0.07$, where the unit is eV.
The energy eigenvalues and eigenvectors of this Hamiltonian are shown in Table B$\cdot$1, where $\ket{\phi_{1}}$, $\ket{\phi_{2}}$, and $\ket{\phi_{3}}$ correspond to sites 1, 2, and 3, respectively.
\begin{table}[htpb]
\caption{Energy eigenvalues and eigenvectors.}
\label{t1}
\begin{center}
\begin{tabular}{ll}
\hline
\multicolumn{1}{c}{energy eigenvalue (eV)} & \multicolumn{1}{c}{eigenvector} \\
\hline
$0.319$ & $-0.769\ket{\phi_{1}}+0.452\ket{\phi_{2}}-0.452\ket{\phi_{3}}$ \\ \hline
$0.042$ & $0.707\ket{\phi_{2}}+0.707\ket{\phi_{3}}$ \\ \hline
$-0.291$ & $-0.639\ket{\phi_{1}}-0.544\ket{\phi_{2}}+0.544\ket{\phi_{3}}$ \\ \hline
\end{tabular}
\end{center}
\end{table}
Since the Fermi energy lies between bands 1 and 2, band 1 is the LUMO band and band 2 is the HOMO band as shown in Fig. B$\cdot$2.
The signs of the eigenvectors and the symmetry of the Wannier orbits determine the parity of the bands. 
Bands 1 and 3 have even parity and band 2 has odd parity. 
The energy difference between the HOMO and LUMO bands is $0.277$ eV, which is almost the same as the energy gap used in the two-orbital model\cite{Kato2020}.

\begin{figure}[htpb]
\begin{center}
\includegraphics[width=50mm]{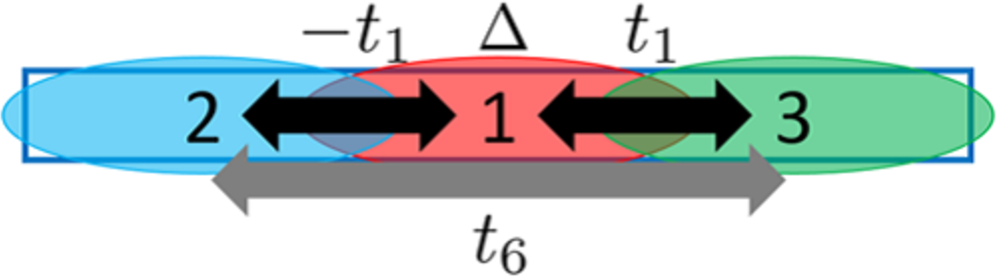}
\end{center}
\caption{(Color online)
One Pt(dmdt)$_{2}$ molecule has three hopping energies. The concrete values of hopping energies are given as $t_{1}=0.212$ and $t_{6}=0.042$. The potential energy of site 1 is $\Delta=0.07$.}
\label{f1}
\end{figure}

\begin{figure}[htpb]
\begin{center}
\includegraphics[width=90mm]{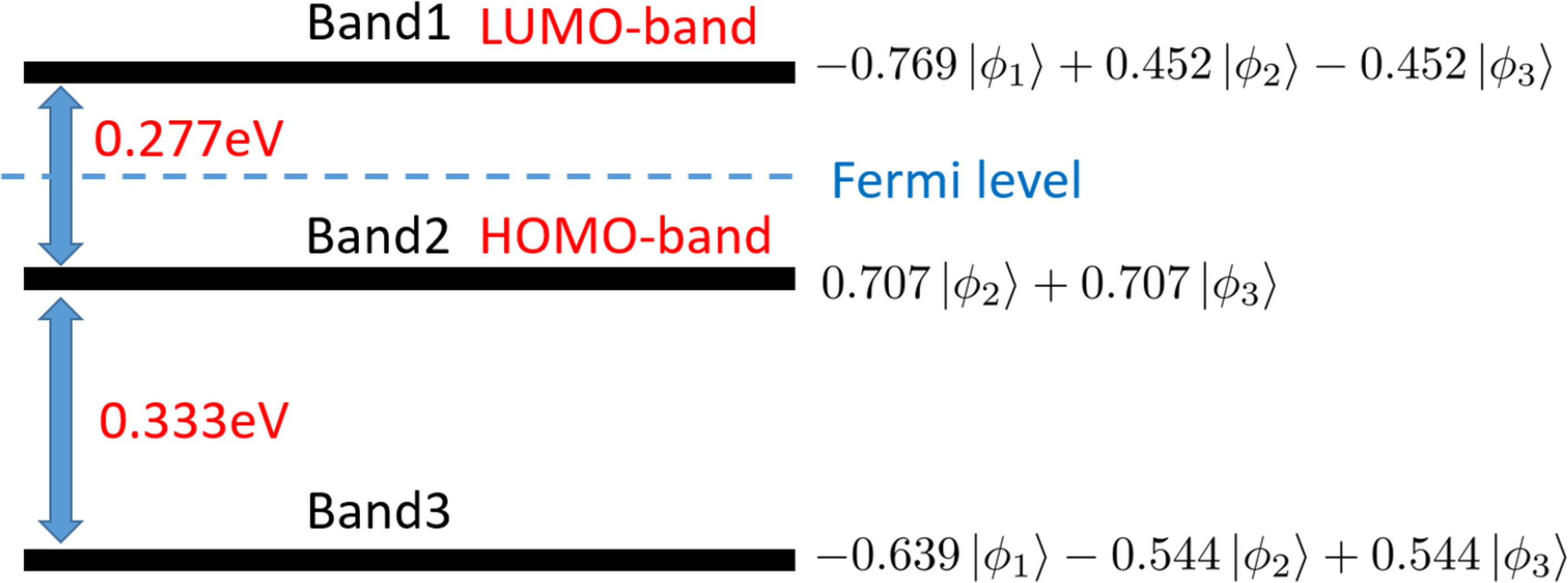}
\end{center}
\caption{(Color online)
Eigenvectors and energy differences between bands 1, 2, and 3. 
The Fermi energy lies between bands 1 and 2. 
The difference between bands 1 (LUMO) and 2 (HOMO) is 0.277 eV.}
\label{f1}
\end{figure}

\end{document}